\documentclass[11pt,a4paper]{article}
\usepackage[hyperref]{acl2020}
\usepackage{times}
\usepackage{latexsym}

\usepackage[utf8]{inputenc} 

\usepackage{mathptmx}
\usepackage{graphicx}
\usepackage{amssymb}
\usepackage{enumerate}
\usepackage{algpseudocode, algorithm}
\usepackage{tabularx}
\usepackage{caption}
\usepackage{subcaption}
\usepackage{booktabs} 
\usepackage{multirow}
\usepackage{tablefootnote}
\usepackage[acronym]{glossaries}
\usepackage{todonotes} 
\usepackage{xspace}
\usepackage{pifont}
\usepackage{microtype}
\newcommand{\xmark}{\ding{55}}%
\newcommand{\cmark}{\ding{51}}%
\usepackage{colortbl}
\usepackage{footmisc}
\usepackage{comment}
\usepackage{natbib}

\newacronym{ld}{LD}{Linked Data}
\newacronym{kg}{KG}{Knowledge Graph}
\newacronym{kb}{KB}{Knowledge Base}

\newcommand{\deer}{\textsc{Deer}\xspace}
\definecolor{missing}{HTML}{FC8D59}
\definecolor{overused}{HTML}{91BFDB}
\newcommand{\hylitemissing}[1]{\cellcolor{missing}\textcolor{black}{#1}}
\newcommand{\hyliteoverused}[1]{\cellcolor{overused}\textcolor{black}{#1}}
\newcommand\doubleplus{+\kern-1ex+\kern0.8ex}

\usepackage{microtype}

\aclfinalcopy 

\title{Instructions for ACL 2020 Proceedings}

\author{Michael R\"oder \and Mohamed Ahmed Sherif \and Muhammad Saleem \\
  \and {\bf Felix Conrads} \and {\bf Axel-Cyrille Ngonga Ngomo} \\
  DICE group, Department of Computer Science\\
  Paderborn University, Germany \\
  Institute for Applied Informatics, Leipzig, Germany} 

\date{}

\title{Benchmarking Knowledge Graphs on the Web}

\begin{document}

\maketitle

\begin{abstract}
The growing interest in making use of Knowledge Graphs for developing explainable artificial intelligence, there is an increasing need for a comparable and repeatable comparison of the performance of Knowledge Graph-based systems. History in computer science has shown that a main driver to scientific advances, and in fact a core element of the scientific method as a whole, is the provision of benchmarks to make progress measurable. 
This paper gives an overview of benchmarks used to evaluate systems that process Knowledge Graphs.
\end{abstract}

\section{Introduction}
\label{42-sec:introduction}

\begin{figure}
\includegraphics[scale=0.20]{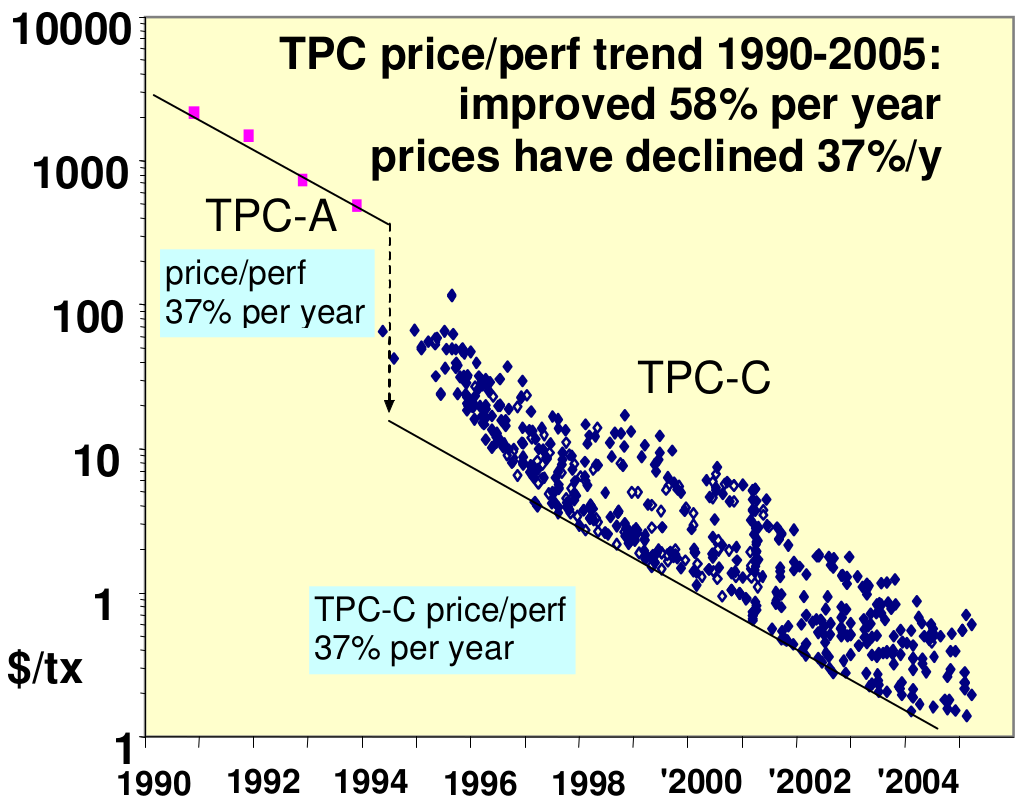}
\caption{Price/performance trend lines for TPC-A and TPC-C. The 15-year trend lines track Moore's Law (100x per 10 years)~\cite{Gray2005}.}
\label{42-fig:tpc-diagram}
\end{figure}

With the growing number of systems using \glspl{kg} there is an increasing need for a comparable and repeatable evaluation of the performance of systems that create, enhance, maintain and give access to \glspl{kg}. 
History in computer science has shown that a main driver to scientific advances,  and in fact a core element of the scientific method as a whole, is the provision of benchmarks to make progress measurable. 
Benchmarks have several purposes: (1) they highlight weak and strong points of systems, (2) they stimulate technical progress and (3) they make technology viable. 
Benchmarks are an essential part of the scientific method as they allow to track the advancements in an area over time and make competing systems comparable. 
The TPC benchmarks\footnote{\url{http://www.tpc.org/information/benchmarks.asp}}  demonstrate how benchmarks can influence an area of research. Even though databases were already quite established in the early 90s, the benchmarking efforts resulted in algorithmic improvements (ignoring advances in hardware) of 15\% per year,  which translated to an order of magnitude improvement in an already very established area~\cite{Gray2005}. 
Figure~\ref{42-fig:tpc-diagram} shows the reduction of the costs per transaction over time.

This paper gives an overview of benchmarks used to evaluate systems that process \glspl{kg}. For creating better comparability between different benchmarks, we determine that each benchmark comprises the following components:
\begin{itemize}
\item The definition of the functionality that will be benchmarked.
\item A set of tasks $T$. Each task $t_j=(i_j, e_j)$ is a pair of input data ($i_j$) and expected output data ($e_j$).
\item Background data $B$ which is typically used to initialize the system.
\item One ore more metrics, which for each task $t_j$ receive the expected result $e_j$, as well as the result $r_j$ provided by the benchmarked system.
\end{itemize}

The remainder of this paper comprises two major parts. First, we present the existing benchmark approaches for \gls{kg}-processing systems. To structure the large list of benchmarks, we follow the steps of the Linked Data lifecycle~\cite{Auer2010,Ngomo2014}. Second, we present existing benchmarking frameworks and briefly summarize their features before concluding the paper.

\section{Benchmarking the Linked Data lifecycle steps}

\begin{figure}
\includegraphics[scale=0.25]{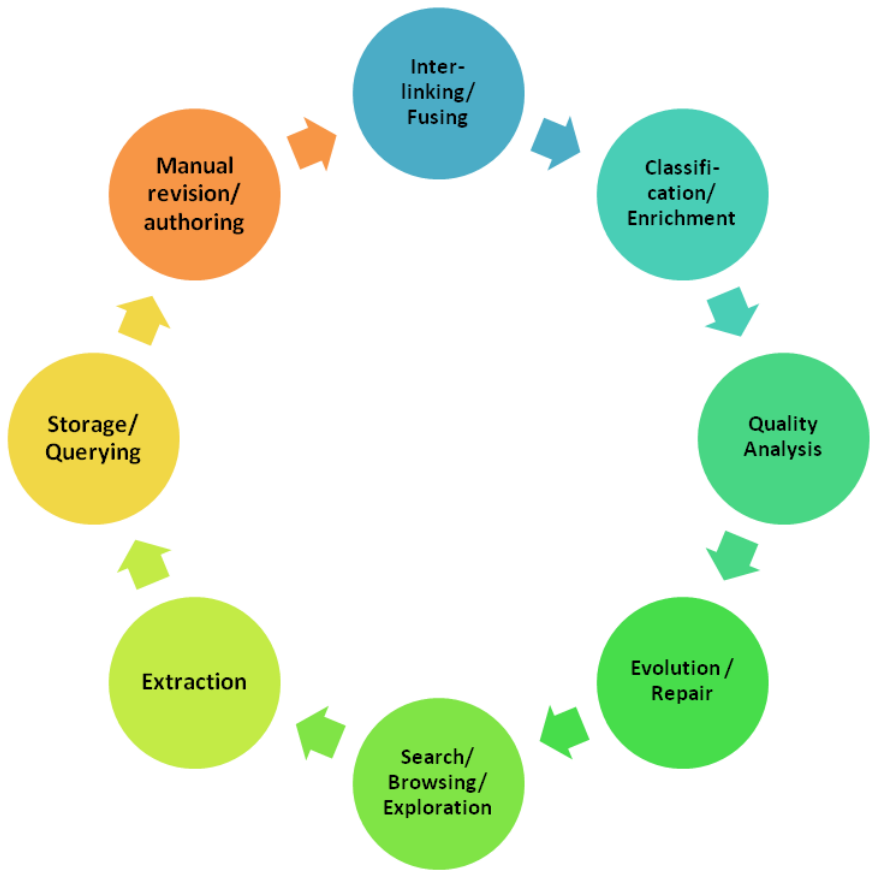}
\caption{The Linked Data lifecycle~\cite{Auer2010,Ngomo2014}.}
\label{42-fig:ldsteps}
\end{figure}

In~\cite{Auer2010,Ngomo2014}, the authors propose a lifecycle of \gls{ld} \glspl{kg} comprising 8 steps. Figure~\ref{42-fig:ldsteps} gives an overview of these steps. In this Section, we go through the single steps, briefly summarize the different actions they cover and how these actions can be benchmarked.

\subsection{Extraction}
\label{42-sec:extraction}

A first step to enter the \gls{ld} lifecycle is to extract information from unstructured or semi-structured representations and transform them into the RDF data model~\cite{Ngomo2014}. This field of information extraction is the subject of research since several decades. The Message Understanding Conference (MUC) introduced a systematic comparison of information extraction approaches in 1993~\cite{Sundheim:1993:TIE:1072017.1072023}. Several other challenges followed, e.g., the Conference on Computational Natural Language Learning (CoNLL) published the CoNLL corpus and organized a shared task on named entity recognition~\cite{conll2003}. 
In a similar way, the Automatic Content Extraction (ACE) challenge~\cite{doddington2004automatic} has been organized by NIST. Other challenges are the Workshop on Knowledge Base Population (TAC-KBP) hosted at the Text Analytics Conference~\cite{mcnamee2009overview} and the Senseval challenge~\cite{kilgarri1998senseval}. In 2013, the Making Sense of Microposts workshop series started including entity recognition and linking challenge tasks~\cite{Cano2013,Cano2014,Rizzo2015,Rizzo2016}. In 2014, the Entity Recognition and Disambiguation (ERD) challenge took place~\cite{ERD2014}. The Open Knowledge Extraction challenge series started in 2015~\cite{oke2015,oke2016,oke2017,oke2018}.

\subsubsection{Extraction Types}
\label{42-tab:extractionTasks}

The extraction of knowledge from unstructered data comes with a wide range of different types. In~\cite{cornolti2013bat}, the authors propose a set of different extraction types that can be benchmarked. These definitions are further refined and extended in~\cite{usbeck2015gerbil,roeder2018gerbil}. A set of basic extraction functionalities for single entities can be distinguished as follows~\cite{roeder2018gerbil}:
\begin{itemize}
\item \emph{ERec}: For this type of extraction the input to the system is a plain text document $d$. The expected output are the mentions $M$ of all entities within the text~\cite{roeder2018gerbil}. A set of entity types $\mathbb T$ is used as background data to define which types of entities should be marked in the texts.
\item \emph{D2KB}: The input for entity disambiguation (or entity linking) is a text $d$ with already marked entity mentions $\mu \in M$. The goal is to map this set of \emph{given} entities to entities from a given \gls{kb} $K$ or to NIL. The latter represents the set of entities that are not present in $K$ (also called emerging entities~\cite{Hoffart:2014:DEE:2566486.2568003}). Although $K$ is not part of the task (and the benchmark datasets do not contain it), it can be seen as background data of a benchmark.
\item \emph{Entity Typing (ET)}: The entity typing is similar to D2KB. Its goal is to map a set of \emph{given} entity mentions to the type hierarchy of a \gls{kb}. 
\item \emph{C2KB}: The concept tagging (C2KB) aims to detect entities relevant for a given document. Formally, the function takes a plain text as input and returns a subset of the \gls{kb}.
\end{itemize}

Based on these basic types, more complex types have been defined~\cite{roeder2018gerbil}:
\begin{itemize}
\item \emph{A2KB}: This is a combination of the ERec and D2KB types and represents the classical named entity recognition and disambiguation. 
Thus, the input is a plain text document $d$ and the system has to identify entities mentions $\mu$ and link them to $K$.
\item \emph{RT2KB}: This extraction type is the combination of entity recognition and typing, i.e., the goal is to identify entities in a given document set $D$ and map them to the types of $K$.
\end{itemize}

For extracting more complex relations the \emph{RE} experiment type is definedin~\cite{speck2018,oke2018}. The input for the relation extraction is a text $d$ with marked entity mentions $\mu \in M$ that have been linked to a knowledge base $K$. Let $u_j \in U$ be the URI of the $j$-th marking $\mu_j$ with $U \subset K \cup \text{NIL}$. The goal is to identify the relation the entities have within the text and return it as triple $(s,p,o)$. Note that the extracted triple does not have to be within the given knowledge base. To narrow the amount of possible relations, a set of properties $\mathbb P$ is given as part of the background data.
Table~\ref{42-tab:extractionTypes} summarizes the different extraction types.

\begin{table*}
\centering
\caption{Summary of the Extraction types.}
\label{42-tab:extractionTypes}
\begin{tabular}{@{}lccc@{}}
\toprule
\multicolumn{1}{c}{\textbf{Type}} & \textbf{Input} $i_j$ & \textbf{Output} $e_j$ & \textbf{Background data} $B$ \\
\midrule
A2KB & $d$ & $\{(\mu,u)|\mu \in M, u \in K \cup \text{NIL}\}$ & $K, \mathbb T$ \\
C2KB & $d$ & $U \subset K$ & $K$ \\
D2KB & $(d,M)$ & $\{(\mu,u)|\mu \in M, u \in K \cup \text{NIL}\}$ & $K$ \\
ERec & $d$ & $M$ & $\mathbb T$ \\
ET   & $(d,M)$ & $\{(\mu,\mathbf{T})|\mu \in M, \mathbf{T} \subset K\}$ & $K$ \\
RE & $(d,M,U)$ & $\{(s,p,o)| s,o \in U, p \in \mathbb P\}$ & $K, \mathbb T$ \\
RT2KB & $d$ & $\{(\mu,c)| c \in K\}$ & $K, \mathbb T$ \\
\bottomrule
\end{tabular}
\end{table*}

\subsubsection{Matchings}
\label{42-sec:Matchings}

One of the major challenges when benchmarking a system is to define how the system's response can be matched to the expected answer. 
This is a non-trivial task when either the expected answer or the system response are sets of elements. A matching defines the conditions that have to be fulfilled by a system response to be a correct result, i.e., to match an element of the expected answer.

\paragraph{Positional Matching.}

Some of the different extraction types described above aim to identify positions of entities within a given text. In the following example, a system recognizes two entities ($e_{s1}$ and $e_{s2}$) that have to be compared to the expected entities ($e_{e1}$ and $e_{e2}$) by the benchmarking framework.

\begin{center}
\begin{tabular}{r}
$e_{s1}$\ \ \ \ \ \ \ \ \ \ \ \ \ \ \ \ \ \ \ \ \ \ \ \ $e_{s2}\qquad$\\
\texttt{|-----|}\ \ \ \ \ \ \ \ \ \ \ \ \ \ \texttt{|-----|}\\
\texttt{President Obama visits Paris.}\\
\texttt{|---------------|}\ \ \ \ \ \ \ \ \ \ \ \ \ \ \texttt{|-----|} \\
$e_{e1}$\ \ \ \ \ \ \ \ \ \ \ \ \ \ \ \ \ \ \ \ \ \ \ \ \ \ \ \ \ \ \ \ \ \ \ $e_{e2}\qquad$\\
\end{tabular}
\end{center}

While the system's marking $e_{s2}$ matches exactly $e_{e2}$, the comparison of $e_{s1}$ and $e_{e1}$ is non-trivial. Two different strategies have been established to handle such cases~\cite{cornolti2013bat,usbeck2015gerbil,roeder2018gerbil}. The \textit{strong annotation matching} matches two markings if they have exactly the same positions. The \textit{weak annotation matching} relaxes this condition. It matches two markings if they overlap with each other.

\paragraph{URI Matching.}

For the evaluation of the system performance with respect to several types of extraction, the matching of URIs is needed. Extraction systems can use different URIs from the creators of a benchmark, e.g., because the extraction system returns Wikipedia article URLs instead of DBpedia URIs. Since different URIs can point to the same real world entity, the matching of two URIs is not as trivial as the comparison of two strings. The authors of~\cite{roeder2018gerbil} propose a workflow for improving the fairness of URI matching:
\begin{enumerate}
\item \emph{URI set retrieval.} Instead of representing a meaning as a single URI, it is represented as a set of URIs comprising URIs that are pointing to the same real world entity. This set can be expanded by crawling the Semantic Web graph using \texttt{owl:sameAs} links and redirects.
\item \emph{URI set classification.} As defined above, the URIs returned by the extraction system either represent an entity that is in $K$ or it represents an emerging entity (NIL). Hence, the after expanding the URI sets, they are classified into these two classes. A URI set $S$ is classified as $S \in C_{KB}$ if it contains at least one URI of $K$. Otherwise it is classified as $S \in C_{EE}$.
\item \emph{URI set matching.} Two URI sets $S_1, S_2$ are matching if both are assigned to the $C_{KB}$ class and the sets are overlapping or both sets are assigned to the $C_{EE}$ class: 
\begin{equation}
\begin{split}
&\left(\left( S_1 \in C_{KB} \right) \land \left( S_2 \in C_{KB} \right)\right. \\
&\quad \left.\land \left( S1 \cap S2 \neq \emptyset \right)\right) \\
\lor & \left(\left( S_1 \in C_{EE} \right) \land \left( S_2 \in C_{EE} \right)\right)
\end{split}
\end{equation}
A comparison of URIs of emerging entities is not suggested since most of these URIs are typically synthetically generated by the extraction system.
\end{enumerate}

\subsubsection{Key Performance Indicators}

The performance of knowledge extraction systems is mainly measured using Precision, Recall and F1-measure. Since most benchmarking datasets comprise multiple tasks, Micro and Macro averages are used for summarizing the performance of the systems~\cite{cornolti2013bat,usbeck2015gerbil,roeder2018gerbil}. A special case is the comparison of entity types. Since the types can be arranged in a hierarchy, the hierarchical F-measure can be used for this kind of experiment~\cite{roeder2018gerbil}. For evaluating the systems' efficiency, the response time of the systems is a typical KPI~\cite{usbeck2015gerbil,roeder2018gerbil}. It should be noted that most benchmarking frameworks send the tasks sequentially to the extraction systems. In contrast to that, the authors of~\cite{oke2017} introduced a stress test. During this test, tasks can be sent in parallel and the gaps between the tasks are reduced over time. For a comparison of the performance of the systems under pressure, a $\beta$ metric is used, which combines the F-measure and the runtime of the system.

\subsubsection{Datasets}
\label{42-sec:extractionDatasets}

\begin{table*}
\centering
\caption{Example benchmarks for knowledge extraction~\cite{roeder2018gerbil}. Collections of datasets, e.g., for a single challenge, have been grouped together.}
\begin{tabular}{@{}lcccrccrccrc@{}}
\toprule
\textbf{Corpus}           & \textbf{Task} & \textbf{Topic} & \multicolumn{3}{c}{$\left|\text{\textbf{Documents}}\right|$} & \multicolumn{3}{c}{\textbf{Entities/Doc.}} & \multicolumn{3}{c}{\textbf{Words/Doc.}} \\
\midrule
ACE2004~\cite{rat:rot}          & A2KB       & news      &&   57 &&&   5.37&&& 373.9 &\\ 
AIDA/CoNLL~\cite{hoffart2011}   & A2KB       & news      && 1393 &&&  25.07&&& 189.7 &\\ 
AQUAINT~\cite{milne2008}        & A2KB       & news      &&   50 &&&  14.54&&& 220.5 &\\ 
Derczynski IPM NEL~\cite{Derczynski}& A2KB       & tweets    &&  182 &&&   1.57&&&  20.8 &\\
ERD2014~\cite{ERD2014}          & A2KB       & queries   &&   91 &&&   0.65&&&   3.5 &\\
GERDAQ~\cite{Cornolti:2016:PSJ:2872427.2883061}           & A2KB       & queries   &&  992 &&&   1.72&&&   3.6 &\\
IITB~\cite{kulkarni2009}        & A2KB       & mixed     &&  103 &&& 109.22&&& 639.7 &\\
KORE 50\tablefootnote{\label{42-foot:yovisto}\url{http://www.yovisto.com/labs/ner-benchmarks/}.}  & A2KB       & mixed     &&   50 &&&   2.88&&&  12.8 &\\ 
Microposts2013~\cite{Cano2013}   & RT2KB      & tweets    && 4265 &&&   1.11&&&  18.8 &\\ 
Microposts2014~\cite{Cano2014}    & A2KB       & tweets    && 3395 &&&   1.50&&&  18.1 &\\
Microposts2015~\cite{Rizzo2015}   & A2KB       & tweets    && 6025 &&&   1.36&&&  16.5 &\\
Microposts2016~\cite{Rizzo2016}   & A2KB       & tweets    && 9289 &&&   1.03&&&  15.7 &\\
MSNBC~\cite{cucerzan2007}         & A2KB       & news      &&   20 &&&  37.35&&& 543.9 &\\ 
N$^3$ Reuters-128~\cite{N3}& A2KB       & news      &&  128 &&&   4.85&&& 123.8 &\\ 
N$^3$ RSS-500~\cite{N3}    & A2KB       & RSS-feeds &&  500 &&&   1.00&&&  31.0 &\\ 
OKE 2015 Task 1~\cite{oke2015}  & A2KB, ET & mixed     &&  199 &&&   5.11&&&  25.5 &\\
OKE 2016 Task 1~\cite{oke2016}  & A2KB, ET  & mixed     &&  254 &&&   5.52&&&  26.6 &\\
Ritter~\cite{Ritter11}           & RT2KB      & news      &&  2394 &&&   0.62&&&  19.4 &\\
Senseval 2~\cite{Edmonds:2001:SO:2387364.2387365}       & ERec       & mixed     &&   242 &&&   9.86&&&  21.3 &\\
Senseval 3~\cite{MihalceaEtAl2004}       & ERec       & mixed     &&   352 &&&   5.70&&&  14.7 &\\
Spotlight Corpus\footref{42-foot:yovisto} & A2KB       & news      &&    58 &&&   5.69&&&  28.6 &\\ 
UMBC~\cite{Finin:2010:ANE:1866696.1866709}             & RT2KB      & tweets     && 12973 &&&   0.97&&&  17.2 &\\
WSDM2012/Meij~\cite{WSDM:2015:blanco}    & C2KB       & tweets    &&   502 &&&   1.87&&&  14.4 &\\
\bottomrule
\end{tabular}
\label{42-tab:extractionDatasets}
\end{table*}

Since the field of information extraction has been tackled for several years, a large number of datasets is available. Table~\ref{42-tab:extractionDatasets} shows some example datasets. It is clear that the datasets vary a lot in their number of documents (i.e., the number of tasks), length of single documents and number of entities per document. All datasets listed in the table have been manually annotated. However, it is known that this type of datasets has two drawbacks. First,~\cite{jha2017eaglet} showed that although the datasets are used as gold standards, they are not free of errors. Apart from wrongly positioned entity mentions, all of these dataset have the issue that the \glspl{kb} used as reference to link the entities have evolved. This leads to annotations that are using outdated URIs not present in the lates \gls{kb} versions. In~\cite{roeder2018gerbil}, the authors propose a mechanism to identify and handle these outdated URIs. 
Second, the size of the datasets is limited by the high costs of the manual annotation of documents. It can be seen from the table that if the number of documents is high, the datasets typically comprises short messages and if the number of words per document is high, the number of documents in the dataset is low. Hence, most of the datasets can not be used to benchmark the scalability of extraction systems. To this end,~\cite{ngomo2018bengal} proposed the automatic generation of annotated documents based on a given \gls{kb}.

\subsection{Storage \& Querying}
\label{42-sec:storage}

After a critical mass of RDF data has been extracted, the data has to be stored, indexed and made available for querying in an efficient way~\cite{Ngomo2014}. Triplestores are data management systems for storing and querying RDF data. In this section, we discuss the different benchmarks used to assess the performance of triplestores. In particular, we highlight key features of triplestore benchmarks pertaining to the three main components of benchmarks, i.e., datasets, queries, and performance metrics. State-of-the-art triplestore benchmarks are analyzed and compared against these features. Most of the content in this section is adopted from \cite{saleem2019representative}.

\subsubsection{Triplestore Benchmark Design Features} 
\label{42-sec:storage-design}
In general, triplestore benchmarks comprise three main components:  (1)~a set of RDF datasets, (2)~a set of SPARQL queries, and (3)~a set of performance metrics. With respect to the benchmark schema defined in Section~\ref{42-sec:introduction}, each task of a SPARQL benchmark comprises a single SPARQL query as input while the expected output data comprises the expected result of the query. The RDF dataset(s) are part of the Background data of a benchmark against which the queries are executed. In the following, we present key features of each of these components that are important to consider in the development of triplestore benchmarks. 

\paragraph{Datasets.} 
\label{42-sec:storage-datasets}
Datasets used in triplestore benchmarks are either synthetic or selected from real-world RDF datasets~\cite{feasible2015}. The use of  real-world RDF datasets is often regarded as useful to perform evaluation in close-to-real-world settings~\cite{dbpsb2011}. Synthetic datasets are useful to test the scalability of systems based on datasets of varying sizes. Synthetic dataset generators are utilized to produce datasets of varying sizes that can often be optimized to reflect the characteristics of real-world datasets~\cite{appleoranges2011}. 
Previous works~\cite{appleoranges2011,rbench2015} highlighted two key measures 
for selecting such datasets for triplestores benchmarking: (1)~Dataset Structuredness and (2)~Relationship Specialty. The formal definitions of these metrics can be found in \cite{saleem2019representative}. However, observations from the literature (see e.g., \cite{largerdfbench2018,appleoranges2011}) suggest that other features such as varying number of triples, number of resources, number of properties, number of objects, number of classes, diversity in literal values, average properties and instances per class, average indegrees and outdegrees as well as their distribution across resources should also be considered. 

\begin{enumerate}
\item \emph{Dataset Structuredness:} Duan et al.~\cite{appleoranges2011} combine many of the aforementioned dataset features into a single composite metric called dataset \emph{structuredness} or \emph{coherence}. This metric measures how well the classes (i.e., \texttt{rdfs:Class}) defined within a dataset are covered by the different instances of this class within the dataset. The structuredness value for any given dataset lies in $[0,1]$, where $0$ stands for the lowest possible structuredness and $1$ represents the highest possible structured dataset. They conclude that synthetic datasets are highly structured while real-world datasets have structuredness values ranging from low to high, covering the whole structuredness spectrum. The formal definition of this metric can be found in \cite{appleoranges2011}.
\item \emph{Relationship Specialty:} In datasets, some attributes are more common and associated with many resources. In addition, some attributes are multi-valued, e.g., a person can have more than one cellphone number or professional skill. The number of occurrences of a predicate associated with each resource in the dataset provides useful information on the graph structure of an RDF dataset, and makes some resources distinguishable from others~\cite{rbench2015}. In real datasets, this kind of relationship specialty is commonplace. For example, several million people can like the same movie. Likewise, a research paper can be cited in several hundred of other publications. Qiao et al.~\cite{rbench2015} suggest that synthetic datasets are limited in how they reflect this relationship specialty. This is either due to the simulation of uniform relationship patterns for all resources, or a random relationship generation process.
\end{enumerate}

The dataset structuredness and relationship specialty directly affect the result size, the number of intermediate results, and the selectivities of the triple patterns of a given SPARQL query. Therefore, they are important dataset design features to be considered during the generation of benchmarks~\cite{appleoranges2011,feasible2015,rbench2015}. 

\paragraph{SPARQL Queries.} 
\label{42-sec:storage-queries}
The literature about SPARQL Queries~\cite{DBLP:conf/semweb/AlucHOD14,splodge2012,feasible2015,largerdfbench2018,sqcframework2017} suggests that a SPARQL querying benchmark should vary the queries with respect to various features such as \emph{query characteristics}: number of triple patterns, number of projection variables, result set sizes, query execution time, number of BGPs, number of join vertices, mean join vertex degree, mean triple pattern selectivities, BGP-restricted and join-restricted triple pattern selectivities, join vertex types, and highly used SPARQL clauses (e.g., \texttt{LIMIT}, \texttt{OPTIONAL}, \texttt{ORDER BY}, \texttt{DISTINCT}, \texttt{UNION}, \texttt{FILTER}, \texttt{REGEX}). All of these features have a direct impact on the runtime performance of triplestores. The formal definitions of many of the above mentioned features can be found in \cite{saleem2019representative}.

\paragraph{Performance Metrics.} 
\label{42-sec:storage-metric}
Based on the previous triplestore benchmarks and performance evaluations \cite{DBLP:conf/semweb/AlucHOD14,fishmark2012,bsbm2009,bowlogna2012,lubm2005,DBLP:conf/sigmod/ErlingALCGPPB15,DBLP:conf/grades/SzarnyasPAMPKEB18,dbpsb2011,feasible2015,sp2bench2009,DBLP:journals/sosym/SzarnyasIRV18,biobenchmark2014,iswc_iguana} the performance metrics for such comparisons can be categorized as: 
\begin{itemize}
    \item \emph{Query Processing Related}: The performance metrics in this category are related to the query processing capabilities of the triplestores. The query execution time is the central performance metric in this category. However, reporting the execution time for individual queries might not be feasible due to the large number of queries in the given benchmark. To this end, Query Mix per Hour (QMpH) and Queries per Second (QpS) are regarded as central performance measures to test the querying capabilities of the triplestores \cite{feasible2015,dbpsb2011,bsbm2009}. In addition, the query processing overhead in terms of the CPU and memory usage is important to measure during the query executions \cite{sp2bench2009}. This also includes the number of intermediate results, the number of disk/memory swaps, etc.
    \item \emph{Data Storage Related:} Triplestores need to load the given RDF data and mostly create indexes before they are ready for query executions. In this regard, the data loading time, the storage space acquired, and the index size are important performance metrics in this category \cite{bowlogna2012,biobenchmark2014,sp2bench2009,bsbm2009}. 
    \item \emph{Result Set Related:} Two systems can only be compared if they produce exactly the same results. Therefore, result set correctness and completeness are important metrics to be considered in the evaluation of triplestores \cite{bsbm2009,sp2bench2009,feasible2015,biobenchmark2014}. 
    \item \textbf{Parallelism with/without Updates:} Some of the aforementioned triplestore performance evaluations \cite{iswc_iguana,biobenchmark2014,bsbm2009} also measured the parallel query processing capabilities of the triplestores by simulating workloads from multiple querying agents with and without dataset updates.
\end{itemize}
We analyzed state-of-the-art existing SPARQL triplestore benchmarks across all of the above mentioned dataset and query features as well as the performance metrics. The results are presented in Section \ref{42-sec:storage-ba}.

\subsubsection{Triplestore Benchmarks}
Triplestore benchmarks can be broadly divided into two main categories, namely synthetic and real-data benchmarks. 

\paragraph{Synthetic Triplestore Benchmarks.}
Synthetic benchmarks make use of the data (and/or query) generators to generate datasets and/or queries for benchmarking. Synthetic benchmarks are useful in testing  the scalability of triplestores with varying dataset sizes and querying workloads. However, such benchmarks can fail to reflect the characteristics of real-world datasets or queries. 
The \emph{Train Benchmark} (TrainBench)~\cite{DBLP:journals/sosym/SzarnyasIRV18} uses a data generator that produces railway networks in increasing sizes and serializes them in different formats, including RDF. \emph{The Waterloo SPARQL Diversity Test Suite} (WatDiv)~\cite{DBLP:conf/semweb/AlucHOD14} provides a synthetic data generator that produces RDF data with a tunable structuredness value and a query generator. The queries are generated from different query templates. 
\emph{SP2Bench}~\cite{sp2bench2009} mirrors vital characteristics (such as power law distributions or Gaussian curves) of the data in the DBLP bibliographic database. 
\emph{The Berlin SPARQL Benchmark} (BSBM)~\cite{bsbm2009} uses query templates to generate any number of SPARQL queries for benchmarking, covering multiple use cases such as explore, update, and business intelligence.
\emph{Bowlogna}~\cite{bowlogna2012} models a real-world setting derived from the Bologna process and offers mostly analytic queries reflecting data-intensive user needs. 
The \emph{LDBC Social Network Benchmark} (SNB) defines two workloads. First, the \emph{Interactive} workload (SNB-INT) measures the evaluation of graph patterns in a localized scope (e.g., in the neighborhood of a person), with the graph being continuously updated~\cite{DBLP:conf/sigmod/ErlingALCGPPB15}.
Second, the \emph{Business Intelligence} workload (SNB-BI) focuses on queries that mix complex graph pattern matching with aggregations, touching on a significant portion of the graph~\cite{DBLP:conf/grades/SzarnyasPAMPKEB18}, without any updates. Note that these two workloads are regarded as two separate triplestore benchmarks based on the same dataset.

\paragraph{Triplestore Benchmarks Using Real Data.}
Real-data benchmarks make use of real-world datasets and queries from real user query logs for benchmarking. Real-data benchmarks are useful in testing triplestores more closely in real-world settings. However, such benchmarks may fail to test the scalability of triplestores with varying dataset sizes and querying workloads. 
\emph{FEASIBLE}~\cite{feasible2015} is a cluster-based SPARQL benchmark generator, which is able to synthesize customizable benchmarks from the query logs of SPARQL endpoints. 
The \emph{DBpedia SPARQL Benchmark} (DBPSB)~\cite{dbpsb2011} is 
another cluster-based approach that generates benchmark queries from DBpedia query logs, but 
employs different clustering techniques than FEASIBLE.
The \emph{FishMark}~\cite{fishmark2012} dataset is obtained from FishBase\footnote{FishBase: \url{http://fishbase.org/search.php}} and provided in both RDF and SQL versions. The SPARQL queries were obtained from logs of the web-based FishBase application. 
\emph{BioBench}~\cite{biobenchmark2014} evaluates the performance of RDF triplestores with biological datasets and queries from five different real-world RDF datasets\footnote{BioBench: \url{http://kiban.dbcls.jp/togordf/wiki/survey\#data}}, i.e., Cell, Allie, PDBJ, DDBJ, and UniProt. Due to the size of the datasets, we were only able to analyze the combined data and queries of the first three.



\autoref{42-tab:highlevel-stats} summarizes the statistics from selected datasets of the benchmarks. More advanced statistics will be presented in the next section. The table also shows the number of SPARQL queries of the datasets included in the corresponding benchmark or query log. It is important to mention that we only selected SPARQL \texttt{SELECT} queries for analysis. This is because we wanted to analyze the triplestore benchmarks for their query runtime performance and most of these benchmarks only contain \texttt{SELECT} queries~\cite{feasible2015}. For the synthetic benchmarks that include data generators, we chose the datasets used in the evaluation of the original paper that were comparable in size to the datasets of other synthetic benchmarks. For template-based query generators such as WatDiv, DBPSB and SNB, we chose one query per available template. For FEASIBLE, we generated a benchmark of 50 queries from the DBpedia log file to be comparable with a well-known WatDiv benchmark that includes 20 basic testing query templates, and 30 extensions for testing.\footnote{The WatDiv query templates are available at \url{http://dsg.uwaterloo.ca/watdiv/}.}

\begin{table*}[!htb]
\setlength{\tabcolsep}{3.5pt}
\centering
\footnotesize
\caption{High-level statistics of the data and queries used on existing triplestore benchmarks. Both SNB-BI and SNB-INT use the same dataset and are therefore named as SNB for simplicity.}
\label{42-tab:highlevel-stats}
\begin{tabular}{@{}clrrrrr@{}}
	\toprule
	                                                                                & Benchmark                                                                       & \multicolumn{1}{c}{Subjects} & \multicolumn{1}{c}{Predicates} & \multicolumn{1}{c}{Objects} & \multicolumn{1}{c}{Triples} & \multicolumn{1}{c}{Queries} \\ \midrule
	\parbox[t]{2mm} {\multirow{6}{*}{\rotatebox[origin=c]{90}{\textbf{Synthetic}}}} & Bowlogna~\cite{bowlogna2012}                                                    &                       2,151k &                        39 &                        260k &                         12M &                    16 \\
	                                                                                & TrainB.~\cite{DBLP:journals/sosym/SzarnyasIRV18}                               &                       3,355k &                        16 &                      3,357k &                         41M &                    11 \\
	                                                                                & BSBM~\cite{bsbm2009}                                                            &                       9,039k &                        40 &                     14,966k &                        100M &                    20 \\
	                                                                                & SP2Bench~\cite{sp2bench2009}                                                    &                       7,002k &                     5,718 &                     19,347k &                         49M &                    14 \\
	                                                                                & WatDiv~\cite{DBLP:conf/semweb/AlucHOD14}                                        &                       5,212k &                        86 &                      9,753k &                        108M &                    50 \\
	                                                                                & SNB~\cite{DBLP:conf/sigmod/ErlingALCGPPB15,DBLP:conf/grades/SzarnyasPAMPKEB18} &                       7,193k &                        40 &                     17,544k &                         46M &                    21 \\ \midrule
	  \parbox[t]{2mm} {\multirow{4}{*}{\rotatebox[origin=c]{90}{\textbf{Real}}}}    & FishMark~\cite{fishmark2012}                                                    &                         395k &                       878 &                      1,148k &                         10M &                    22 \\
	                                                                                & BioBench~\cite{biobenchmark2014}                                                &                     278,007k &                       299 &                    232,041k &                      1,451M &                    39 \\
	                                                                                & FEASIBLE~\cite{feasible2015}                                                    &                      18,425k &                    39,672 &                     65,184k &                        232M &                    50 \\
	                                                                                & DBPSB~\cite{dbpsb2011}                                                          &                      18,425k &                    39,672 &                     65,184k &                        232M &                    25 \\ 
 \bottomrule
\end{tabular}
\end{table*}

\subsubsection{Analysis of the Triplestore Benchmarks}
\label{42-sec:storage-ba} 
We present a detailed analysis of the datasets, queries, and performance metrics of the selected benchmarks and datasets according to the design features presented in Section \ref{42-sec:storage-design}.

\paragraph{Datasets.}
We presents results pertaining to the dataset features of Section~\ref{42-sec:storage-datasets}.

\begin{figure*}
\centering
\begin{subfigure}{.50\textwidth}
  \centering
  \includegraphics[width=\textwidth]{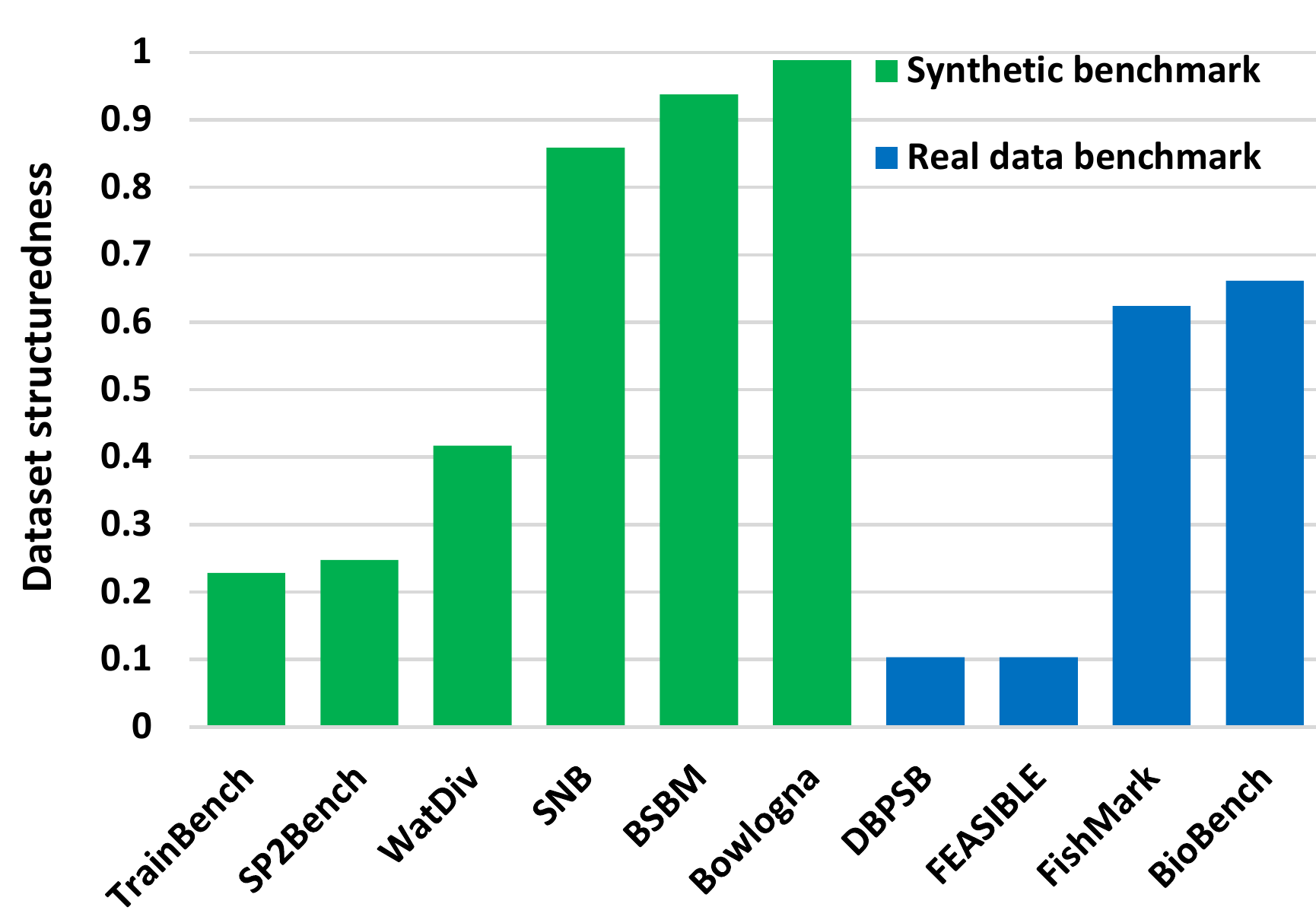}
  \caption{Structuredness}
  \label{42-fig:structuredness}
\end{subfigure}%
\begin{subfigure}{.50\textwidth}
  \centering
  \includegraphics[width=\textwidth]{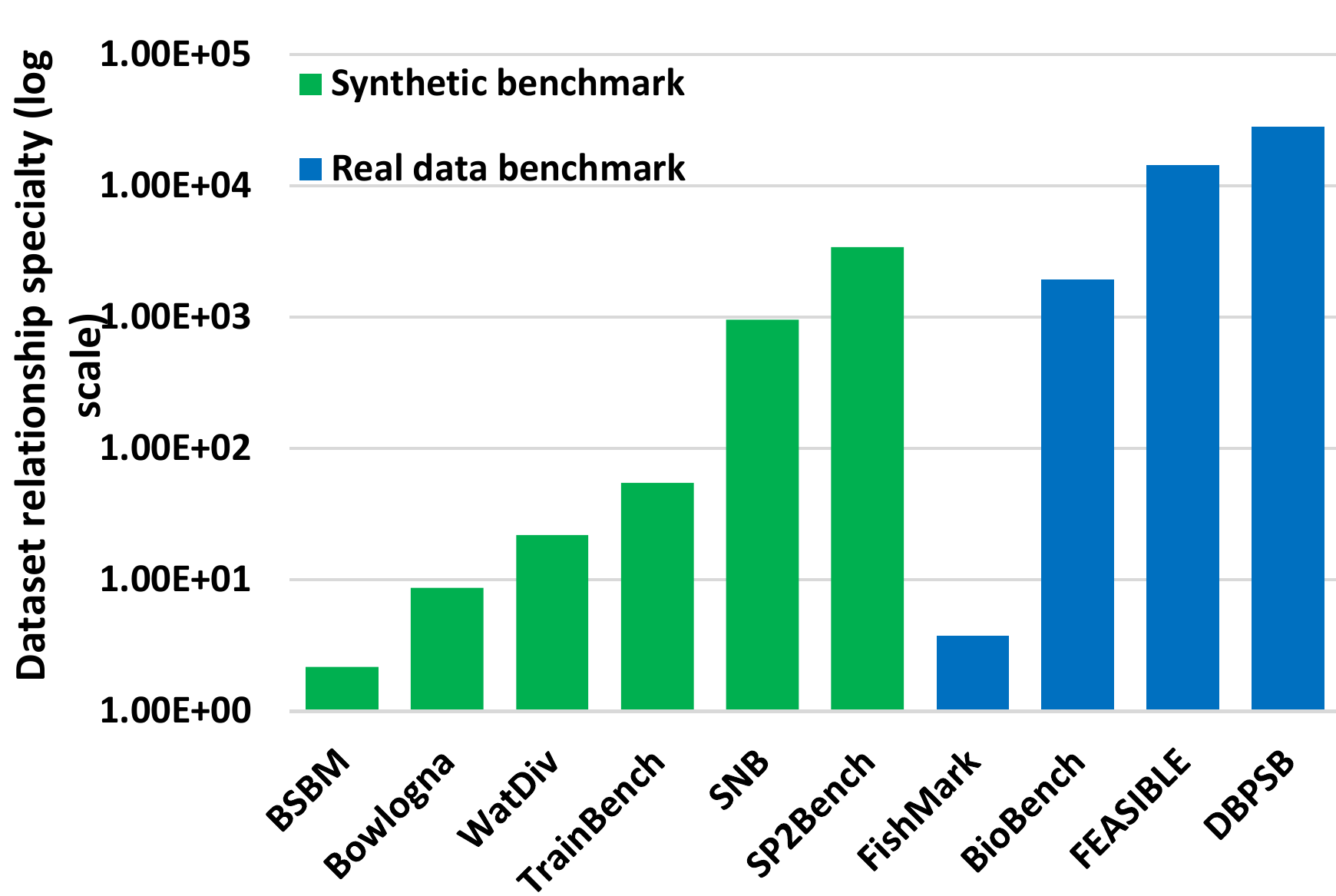}
  \caption{Relationship Specialty}
  \label{42-fig:specialty}
\end{subfigure}
\caption{Analysis of the datasets of triplestore benchmarks and real-world data.}
\label{42-fig:fdvsbb}
\end{figure*}

\begin{itemize}
\item \emph{Structuredness.} Figure~\ref{42-fig:structuredness} shows the structuredness values of the selected benchmarks. Duan et al.~\cite{appleoranges2011} establish that synthetic benchmarks are highly structured while real-world datasets are low structured. This important dataset feature is well-covered in recent synthetic benchmarks such as TrainBench (with a structuredness value of 0.23) and WatDiv, which lets the user generate a benchmark dataset of a desired structuredness value. However, Bowlogna (0.99), BSBM (0.94), and SNB (0.86) have relatively high structuredness values. Finally, on average, synthetic benchmarks are still more structured than real data benchmarks (0.61 vs. 0.45).
\item \emph{Relationship Specialty.} According to~\cite{rbench2015}, relationship specialty in synthetic datasets is limited, i.e., the overall relationship specialty values of synthetic datasets are lower 
than those of similar real-world datasets. The dataset relationship specialty results presented in Figure~\ref{42-fig:specialty} mostly confirm this behavior. On average, synthetic benchmarks have a smaller specialty score than real-world datasets. The relationship specialty values of Bowlogna (8.7), BSBM (2.2), and WatDiv (22.0) are on the lower side compared to real-world datasets. The highest specialty value (28282.8) is recorded in the DBpedia 
dataset. 
\end{itemize}

An important issue 
is the correlation between structuredness and the relationship specialty of the datasets. To this end, we computed the Spearman's rank correlation between the stucturedness and specialty values of all the selected benchmarks. The correlation of the two measures is $-0.5$, indicating a moderate inverse relationship. This means that the higher the structuredness, the lower the specialty value. This is because in highly structured datasets, data is generated according to a specific distribution without treating some predicates more particularly (in terms of occurrences) than others.

\paragraph{Queries.}
This section presents results pertaining to the query features discussed in Section \ref{42-sec:storage-queries}. 
Figure~\ref{42-fig:query-analysis} shows the box plot 
distributions of real-world datasets and benchmark queries across the query features defined in Section \ref{42-sec:storage-design}. The values inside the brackets, e.g., the 0.89 in ``BioBench (0.89)'', show the diversity score of the benchmark for the given query feature. 

\begin{figure*}
\centering
\begin{subfigure}{.50\textwidth}
  \centering
  \includegraphics[scale=0.3]{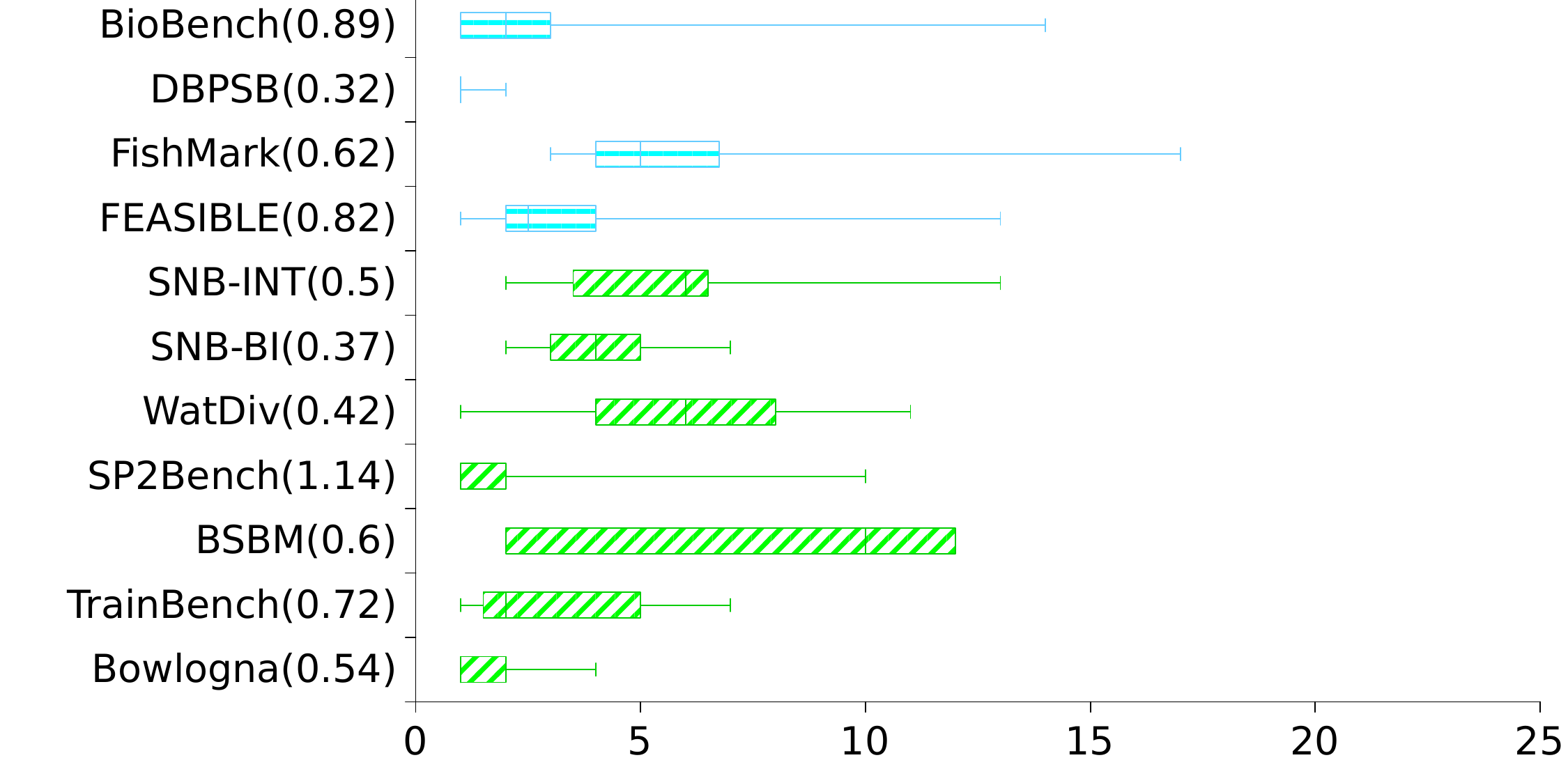}
  \caption{No. of projection variables}
  \label{42-fig:pv}
\end{subfigure}%
\begin{subfigure}{.50\textwidth}
  \centering
  \includegraphics[scale=0.3]{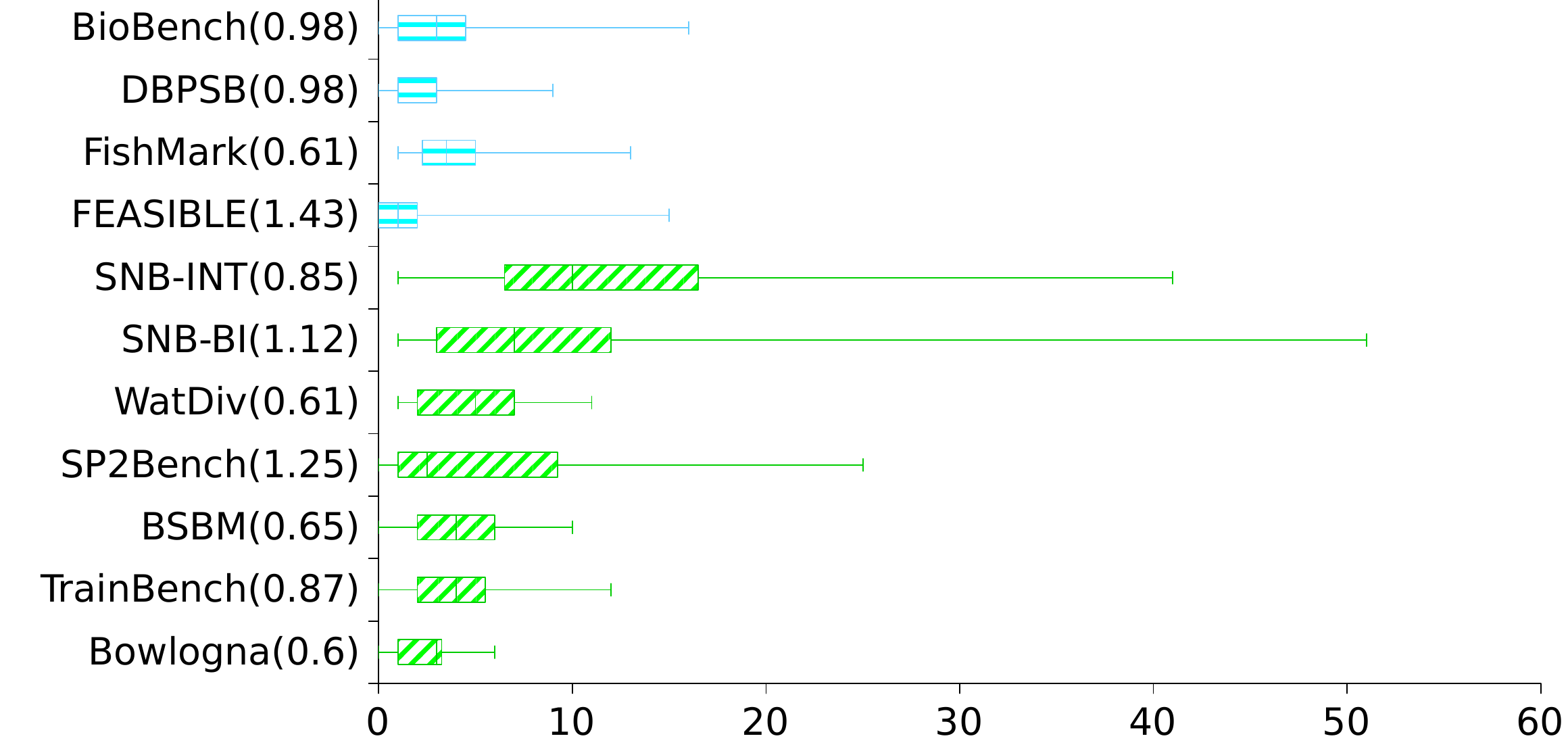}
  \caption{No. of join vertices}
  \label{42-fig:jv}
\end{subfigure}
\begin{subfigure}{.50\textwidth}
  \centering
  \includegraphics[scale=0.3]{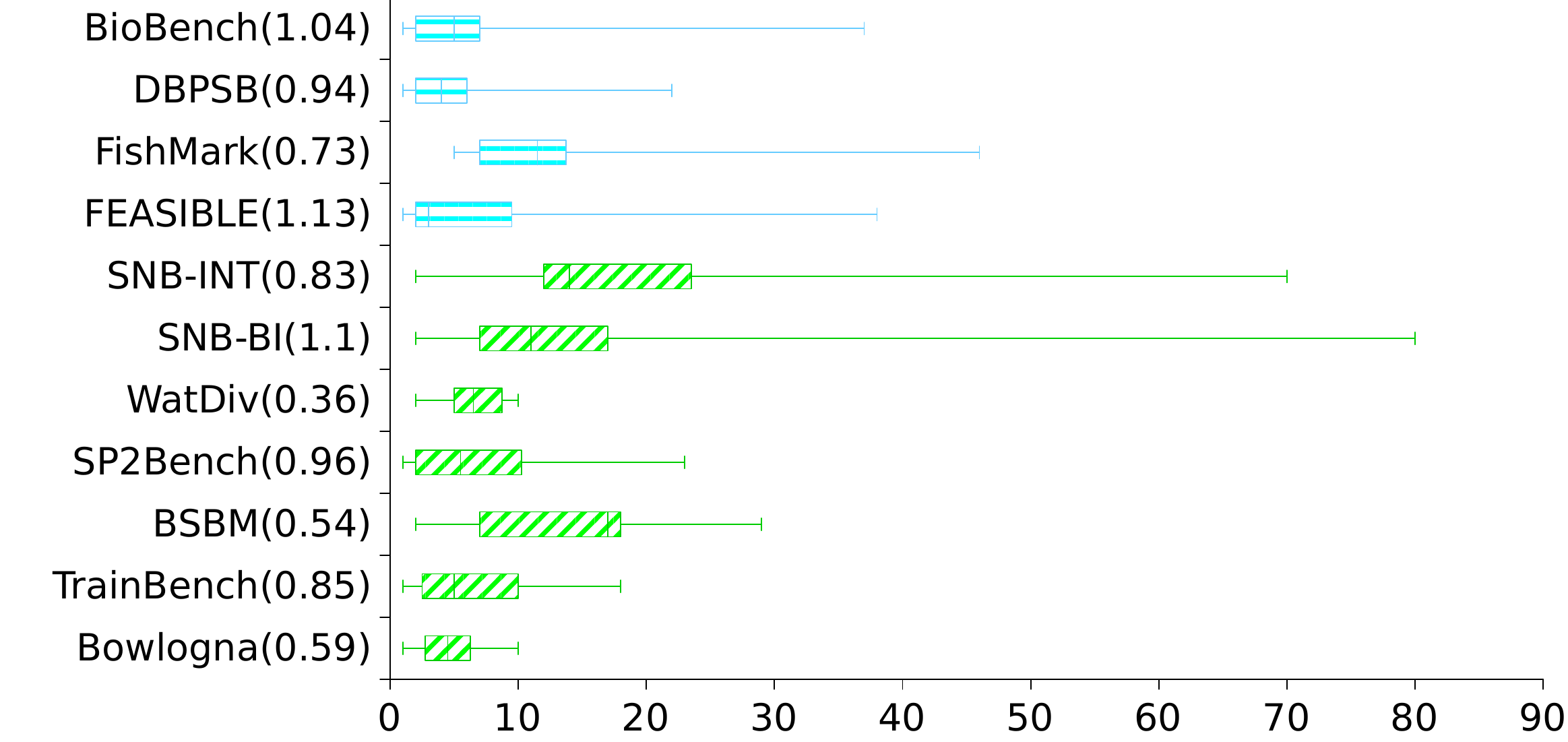}
  \caption{No. of triple patterns}
  \label{42-fig:tp}
\end{subfigure}%
\begin{subfigure}{.50\textwidth}
  \centering
  \includegraphics[scale=0.3]{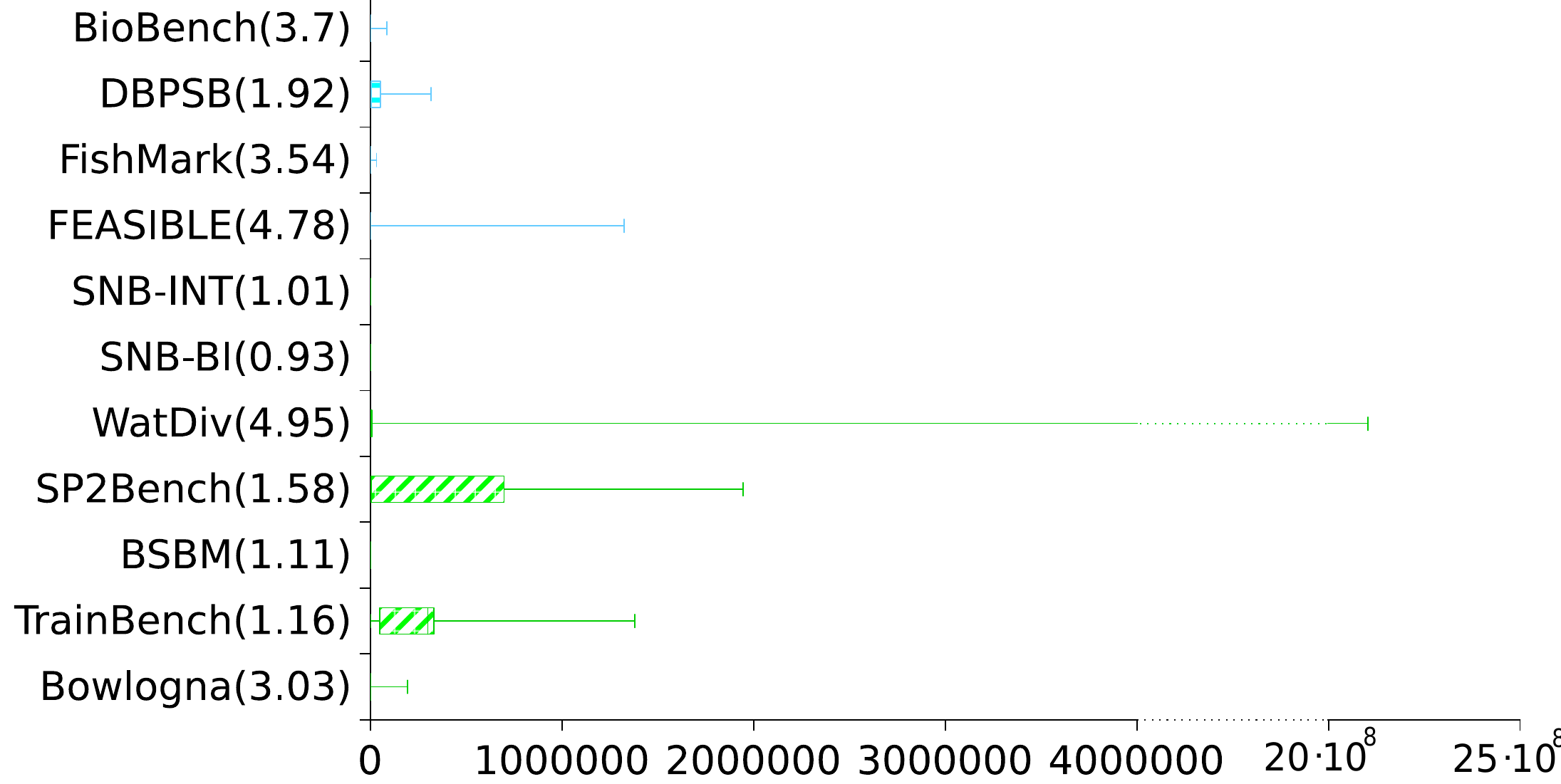}
  \caption{Result size}
  \label{42-fig:rs}
\end{subfigure}
\begin{subfigure}{.50\textwidth}
  \centering
  \includegraphics[scale=0.3]{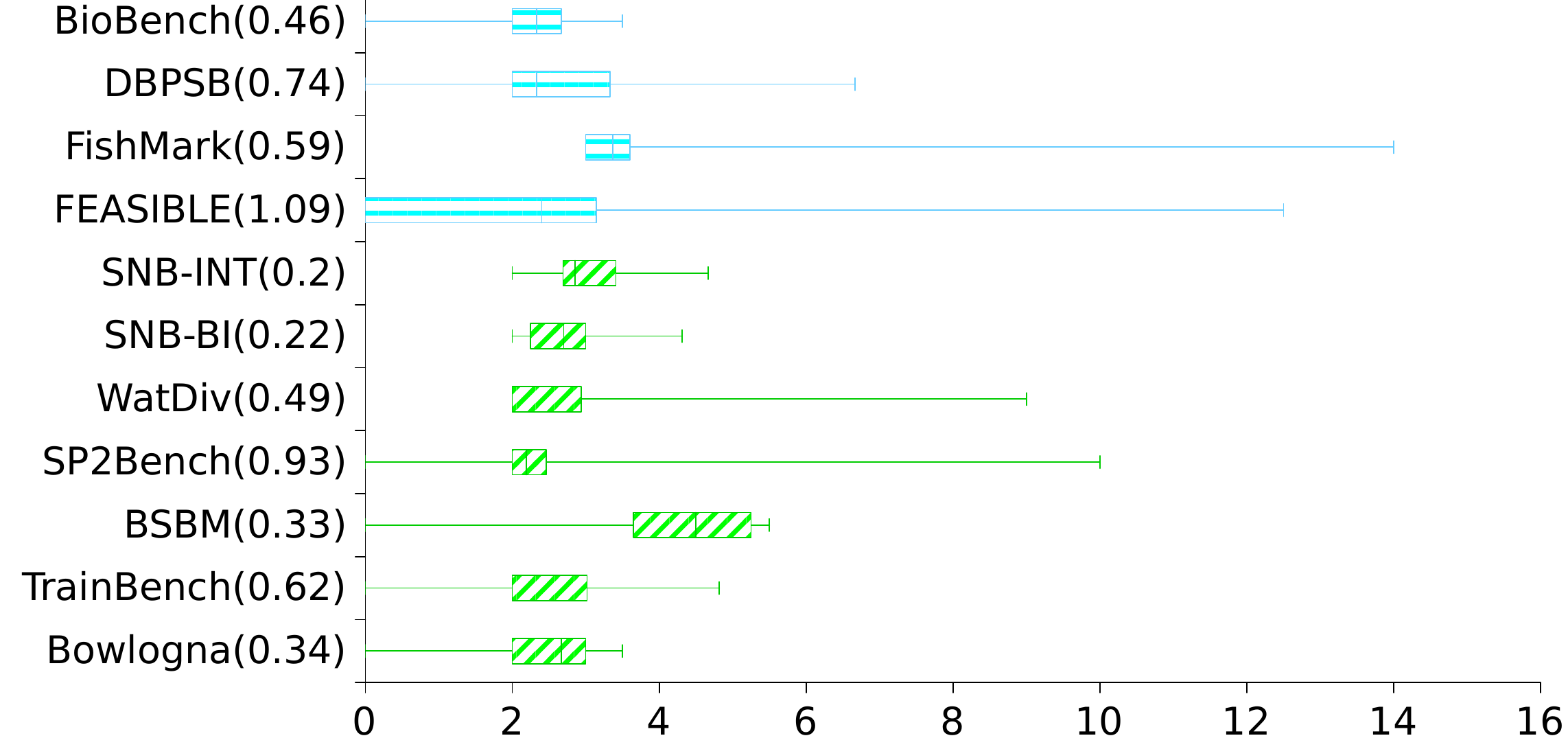}
  \caption{Join vertex degree}
  \label{42-fig:jvd}
\end{subfigure}%
\begin{subfigure}{.50\textwidth}
  \centering
  \includegraphics[scale=0.3]{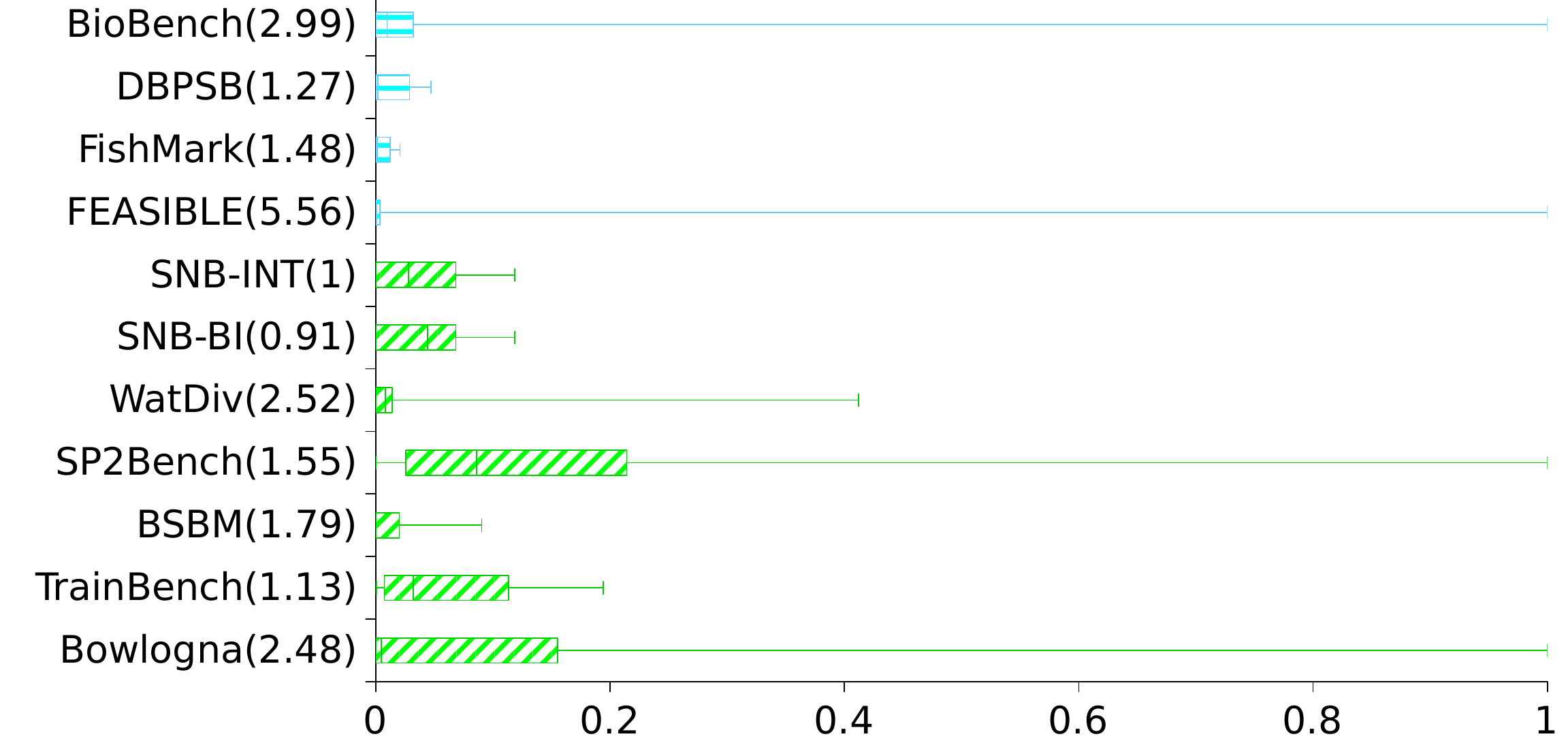}
  \caption{Triple pattern (TP) selectivity}
  \label{42-fig:tpsel}
\end{subfigure}
\begin{subfigure}{.50\textwidth}
  \centering
  \includegraphics[scale=0.3]{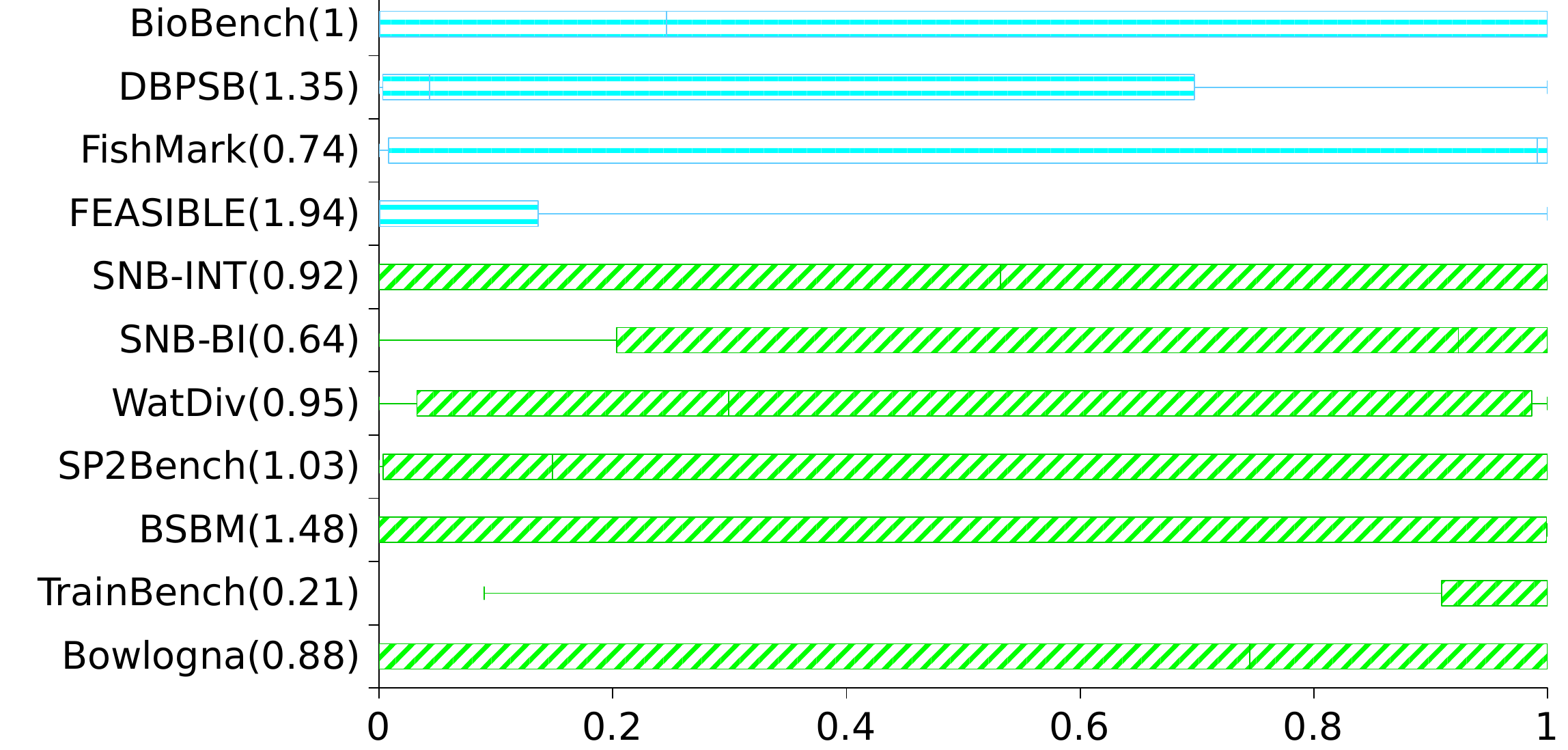}
  \caption{Join-restricted TP selectivity}
  \label{42-fig:jtpsel}
\end{subfigure}%
\begin{subfigure}{.50\textwidth}
  \centering
  \includegraphics[scale=0.3]{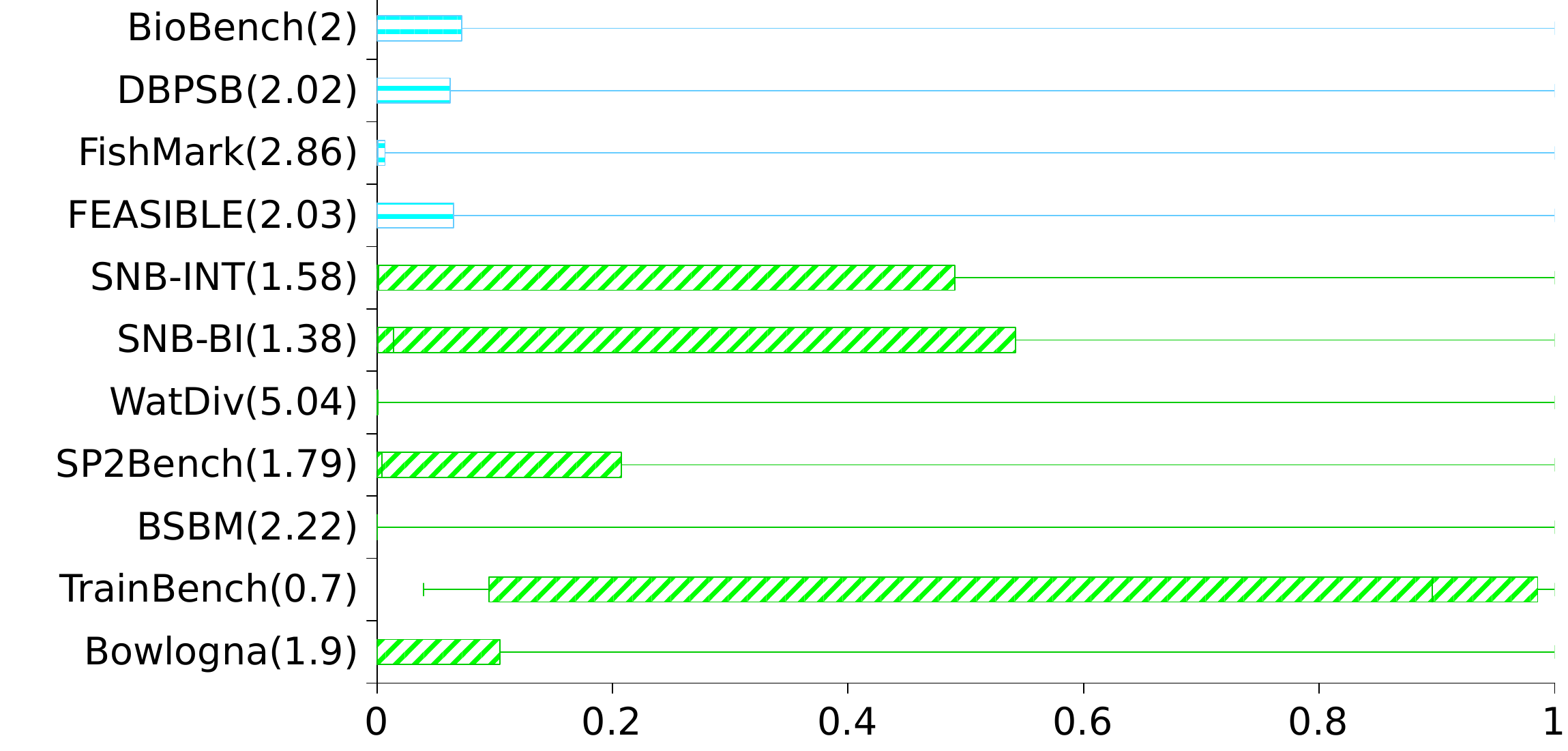}
  \caption{BGP-restricted TP selectivity}
  \label{42-fig:btpsel}
\end{subfigure}
\begin{subfigure}{.50\textwidth}
  \centering
  \includegraphics[scale=0.3]{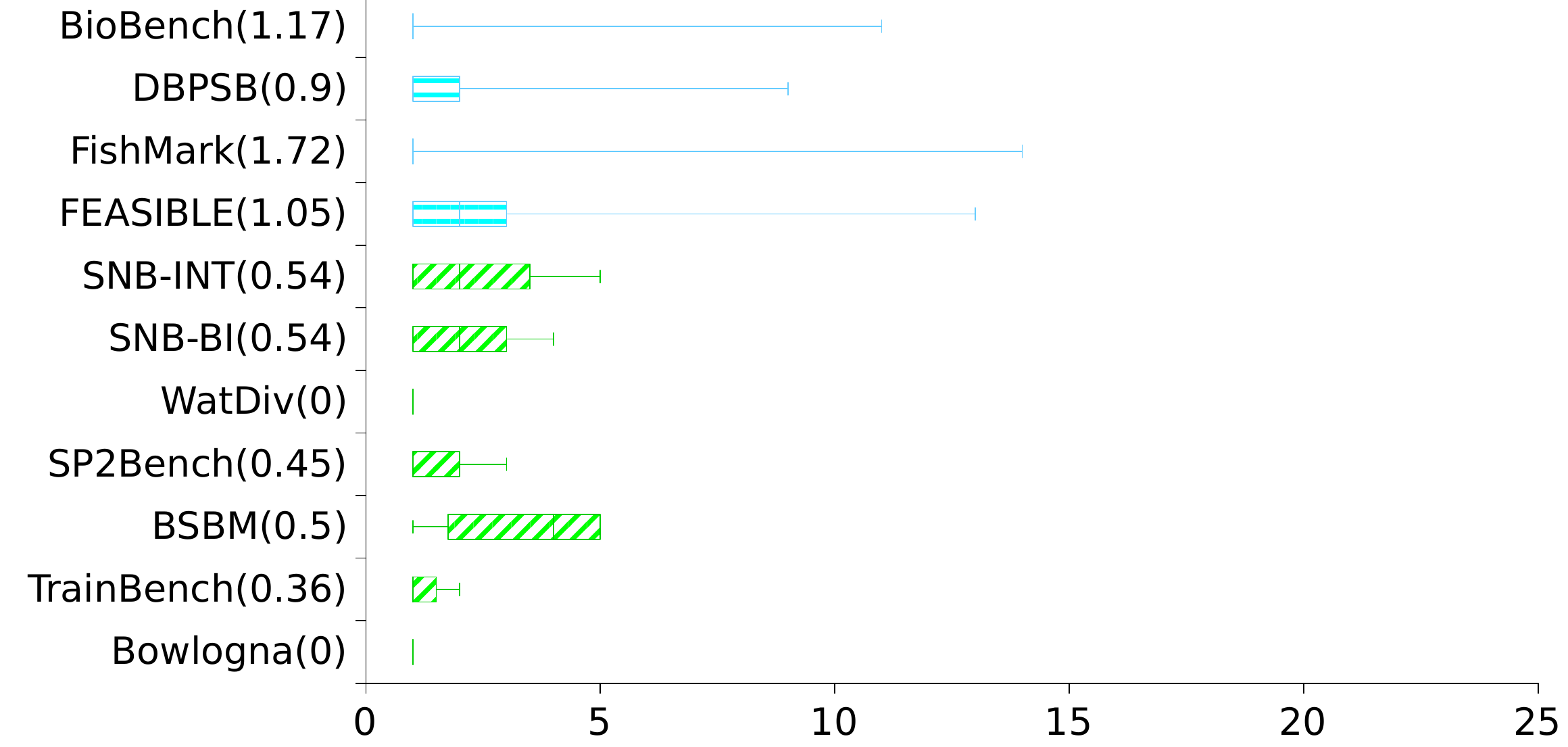}
  \caption{No. of BGPs}
  \label{42-fig:bgps}
\end{subfigure}%
\begin{subfigure}{.50\textwidth}
  \centering
  \includegraphics[scale=0.3]{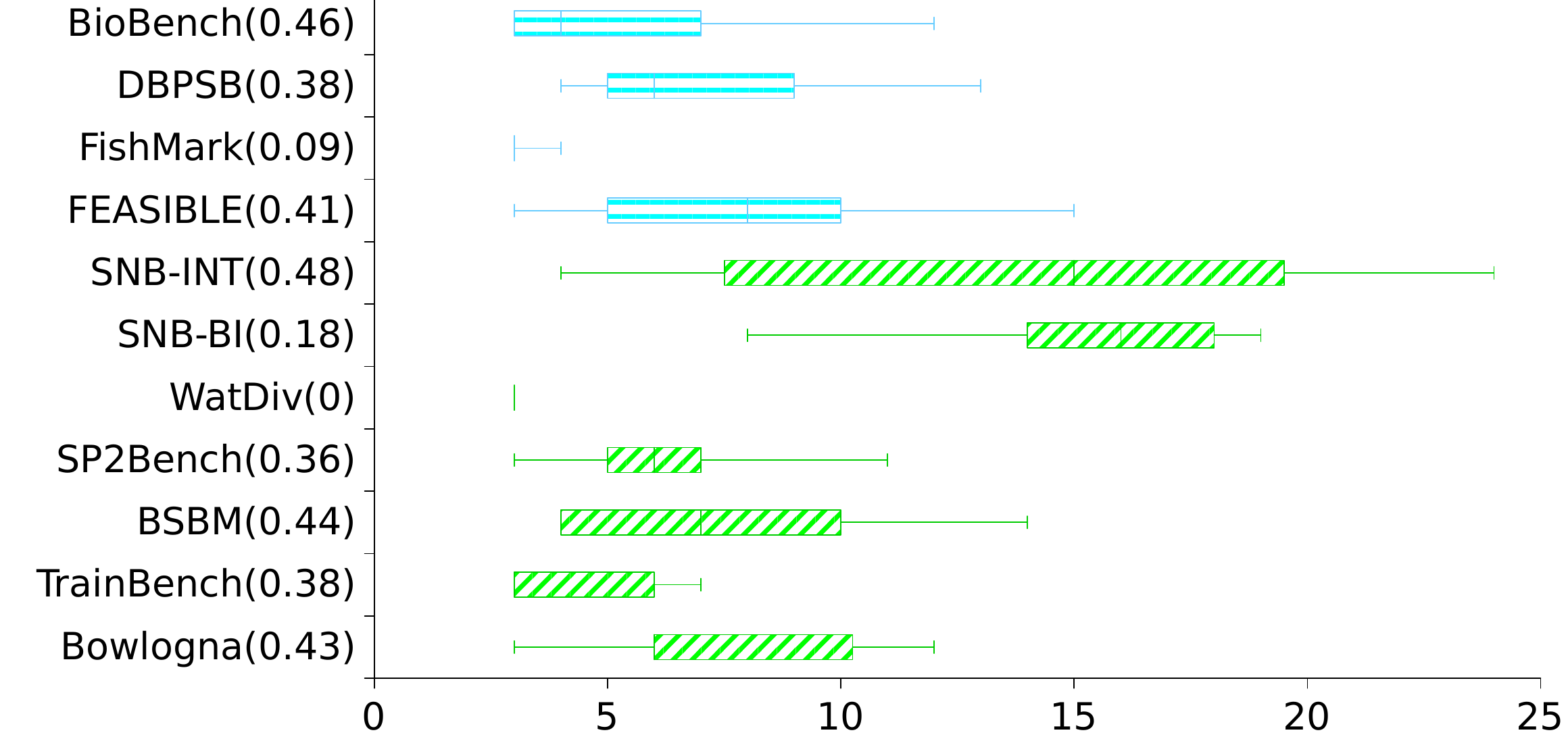}
  \caption{No. of LSQ features}
  \label{42-fig:nlsq}
\end{subfigure}
\label{42-fig:fdvsbb1}
\end{figure*}

\begin{figure*}[!htb]
\ContinuedFloat
\centering

\begin{subfigure}[b]{.50\textwidth}
  \centering
  \includegraphics[scale=0.3]{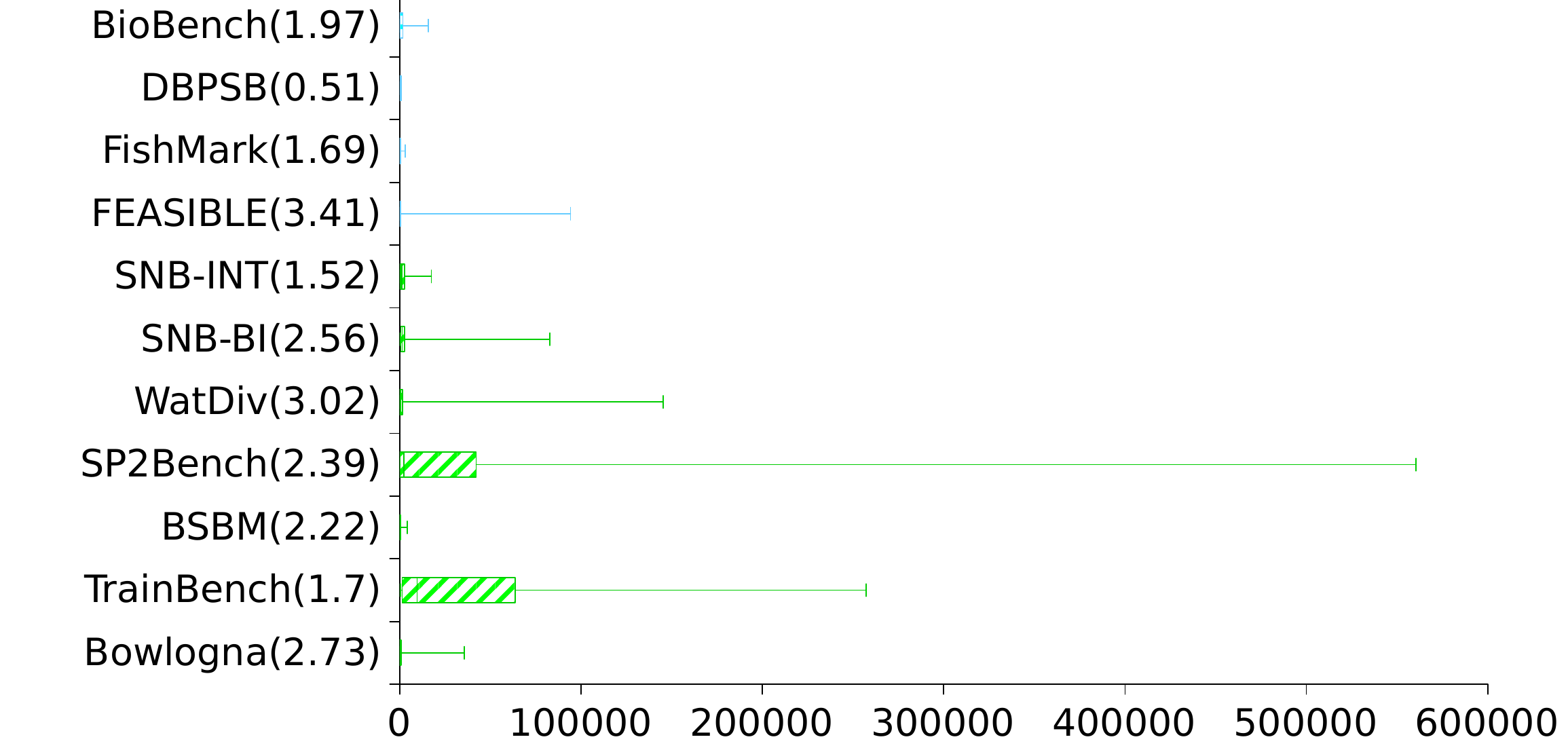}
  \caption{Runtimes (ms)}
  \label{42-fig:rt}
\end{subfigure}%
\begin{subfigure}[b]{.50\textwidth}
  \centering
  \includegraphics[width=\textwidth]{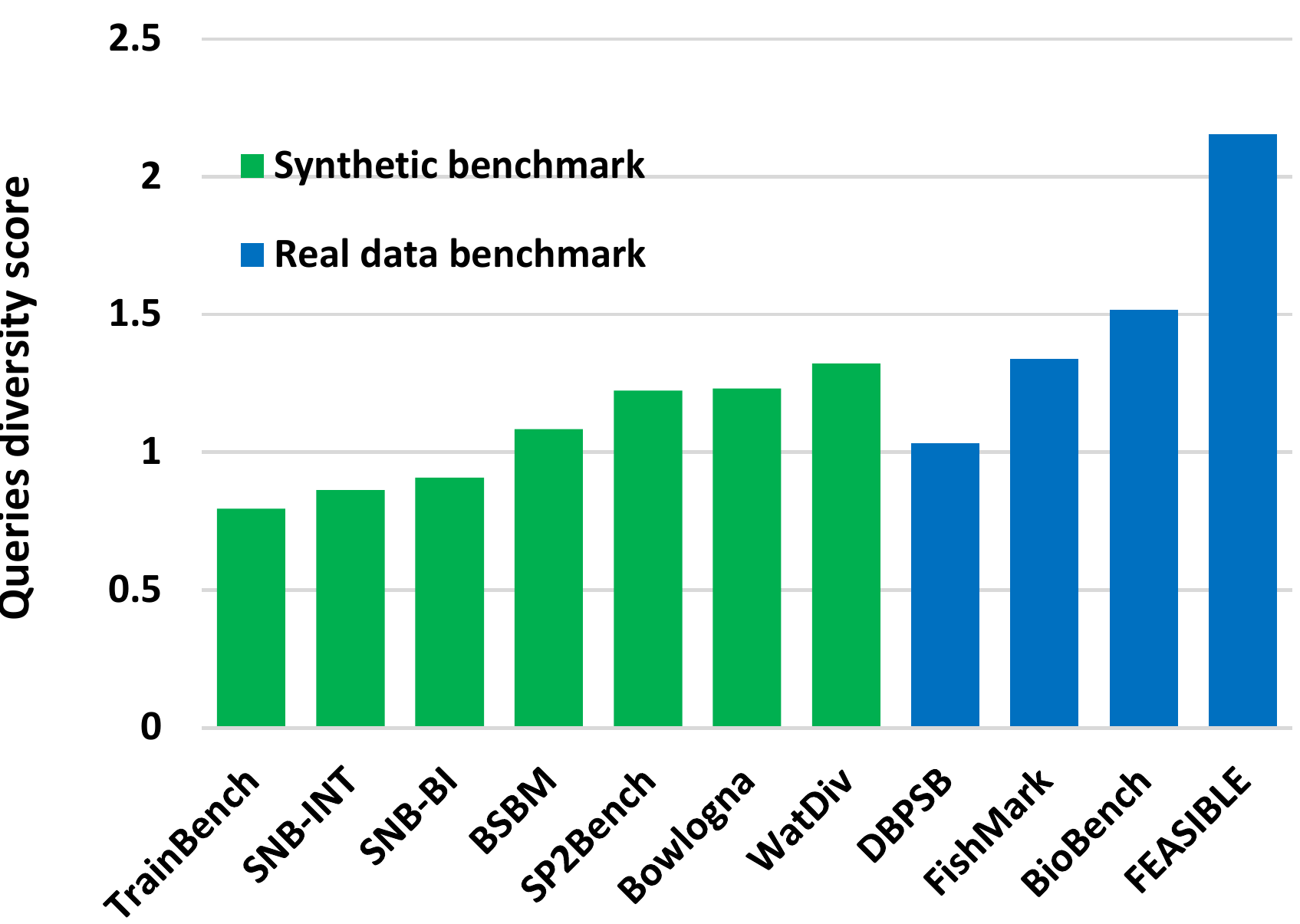}
  \caption{Overall diversity score}
  \label{42-fig:odiversity}
\end{subfigure}
\caption{Analysis of queries used in triplestore benchmarks and for real-world datasets.}
\label{42-fig:query-analysis}
\vspace{-1.9ex}
\end{figure*}

Starting from the number of projection variables (ref. Figure~\ref{42-fig:pv}), the DBPSB dataset has the lowest diversity score (0.32) and SP2Bench has the highest score (1.14). The diversity scores of DBPSB, SNB-BI, SNB-INT, WatDiv, BSBM, WatDiv, and Bowlogna are below the average value (0.63). 
The average diversity score of the number of join vertices (ref. Figure~\ref{42-fig:jv}) is 0.91 and hence the diversity scores of the Bowlogna, FishMark, WatDiv, BSBM, SNB-INT, and TrainBench are below the average value. It is important to mention that the highest number of join vertices recorded in a query is 51 in the SNB-BI benchmark.
The average diversity score of the number of triple patterns (ref. Figure~\ref{42-fig:tp}) is 0.83 and hence the diversity scores of the FishMark, Bowlogna, BSBM, and WatDiv benchmarks are below the average value. 
The average diversity score of the result sizes (ref. Figure~\ref{42-fig:rs}) is 2.29 and hence the diversity scores of SNB-BI, SNB-INT, BSBM, TrainBench, SP2Bench, and DBPSB are below the average value. 
The average diversity score of the join vertex degree (ref. Figure~\ref{42-fig:jvd}) is 0.55 and hence the diversity scores of SNB-BI, SNB-INT, BSBM, Bowlogna, BioBench, and WatDiv are below the average value. 
The average diversity score of the triple pattern selectivity (ref. Figure~\ref{42-fig:tpsel}) is 2.06 and hence the diversity scores of SNB-BI, SNB-INT, TrainBench, DBPSB, FishMark, SP2Bench, and BSBM are below the average. 
The average diversity score of the join-restricted triple pattern selectivity (ref. Figure~\ref{42-fig:jtpsel}) is 1.01 and hence the diversity scores of TrainBench, SNB-BI, SNB-INT, FishMark, Bowlogna, WatDiv, and BioBench are below the average value. 
The average diversity score of the BGP-restricted triple pattern selectivity (ref. Figure~\ref{42-fig:btpsel}) is 2.14 and hence the diversity scores of TrainBench, SNB-BI, SNB-INT, SP2Bench, Bowlogna
BioBench, DBPSB, and FEASIBLE are below the average value. 
The average diversity score of the number of BGPs (ref. Figure~\ref{42-fig:bgps}) is 0.66 and hence the diversity scores of SNB-BI, SNB-INT, BSBM, SP2BENCH, TrainBench, WatDiv, and Bowlogna are below the average value. 

The Linked SPARQL Queries (LSQ)~\cite{lsq2015} representation stores additional SPARQL features, such as the use of \texttt{DISTINCT}, \texttt{REGEX}, 
\texttt{BIND}, \texttt{VALUES}, 
\texttt{HAVING}, \texttt{GROUP BY}, \texttt{OFFSET}, aggregate functions, \texttt{SERVICE}, \texttt{OPTIONAL}, \texttt{UNION},
property paths, etc. We make a count of all of these SPARQL operators and functions and use it as a single query dimension as number of LSQ features. The average diversity score of the number of LSQ features (ref. Figure~\ref{42-fig:nlsq}) is 0.33, and hence only the diversity scores of SNB and WatDiv are below average value. 
Finally, the average diversity score of the query runtimes is 2.04 (ref. Figure~\ref{42-fig:rt}), and hence the diversity scores of DBPSB, FishMark, TrainBench, WatDiv, and BioBench  are below average value.    

In summary, FEASIBLE generates the most diverse benchmarks, followed by BioBench, FishMark, WatDiv, BowLogna, SP2Bench, BSBM, DBPSB, SNB-BI, SNB-INT, and TrainBench. 

\autoref{42-tab:coverages} shows the  percentage coverage of widely used~\cite{lsq2015} SPARQL clauses and join vertex types for each benchmark and real-world dataset. We highlighted cells for benchmarks that either completely miss or overuse certain SPARQL clauses and join vertex types. TrainBench and WatDiv queries mostly miss the important SPARQL clauses. All of FishMark's queries contain at least one ``Star" join node. The distribution of other SPARQL clauses, such as subquery, \texttt{BIND}, aggregates, solution modifiers, property paths, and services are provided in the LSQ versions of each of the benchmark queries, available from the project website. 

\begin{table*}[!htb]
	\centering
	\footnotesize
	\caption{Coverage of SPARQL clauses and join vertex types for each benchmark in percentages. 
		SPARQL clauses:
		\texttt{DIST[INCT]},
		\texttt{FILT[ER]},
		\texttt{REG[EX]},
		\texttt{OPT[IONAL]},
		\texttt{UN[ION]},
		\texttt{LIM[IT]},
		\texttt{ORD[ER BY]}.
		Join vertex types:
		Star, Path, Sink, Hyb[rid], N[o] J[oin].
		Missing \textcolor{missing}{$\blacksquare$} 
		and	overused \textcolor{overused}{$\blacksquare$} 
		features are highlighted.
	}
	\label{42-tab:coverages}
	\setlength\tabcolsep{4.4pt}
	\begin{tabular}{@{}clrrrrrrrrrrrr@{}}
		\toprule
&& \multicolumn{7}{c}{Distributions of SPARQL Clauses}                                                                                & \multicolumn{5}{c}{Distr.\ of Join Vertex Type}                                           \\
\cmidrule(lr){3-9} \cmidrule(l){10-14}
& Benchmark 
& \multicolumn{1}{c}{\tt DIST} 
& \multicolumn{1}{c}{\tt FILT} 
& \multicolumn{1}{c}{\tt REG} 
& \multicolumn{1}{c}{\tt OPT} 
& \multicolumn{1}{c}{\tt UN} 
& \multicolumn{1}{c}{\tt LIM} 
& \multicolumn{1}{c}{\tt ORD} 
& \multicolumn{1}{c}{Star} 
& \multicolumn{1}{c}{Path} 
& \multicolumn{1}{c}{Sink} 
& \multicolumn{1}{c}{Hyb.} 
& \multicolumn{1}{c}{N.J.} \\ \midrule
\parbox[t]{2mm} {\multirow{6}{*}{\rotatebox[origin=c]{90}{\textbf{Synthetic}}}} 
& Bowlogna  
& 6.2 
& 37.5 
& 6.2 
& \hylitemissing{0.0} 
& \hylitemissing{0.0} 
& 6.2 
& 6.2 
& 93.7 
& 37.5 
& 62.5 
& 25.0 
& 6.2 \\
& TrainB.  
& \hylitemissing{0.0} 
& 45.4 
& \hylitemissing{0.0} 
& \hylitemissing{0.0} 
& \hylitemissing{0.0} 
& \hylitemissing{0.0} 
& \hylitemissing{0.0}  
& 81.8 
& 27.2 
& 72.7 
& 45.4 
& 18.1 \\
& BSBM      
&                         30.0 
&                         65.0 
&         \hylitemissing{0.0} 
&                        65.0 
&                       10.0 
&                        45.0 
&                        45.0 
&                     95.0 
&                     60.0 
&                     75.0 
&                     60.0 
&                    5.0 \\
		                                                                                & SP2Bench  &                         42.8 &                         57.1 &         \hylitemissing{0.0} &                        21.4 &                       14.2 &                         7.1 &                        14.2 &                     78.5 &                     35.7 &                     50.0 &                     28.5 &                   14.2 \\
		                                                                                & Watdiv    &          \hylitemissing{0.0} &          \hylitemissing{0.0} &         \hylitemissing{0.0} &         \hylitemissing{0.0} &        \hylitemissing{0.0} &         \hylitemissing{0.0} &         \hylitemissing{0.0} &                     28.0 &                     64.0 &                     26.0 &                     20.0 &    \hylitemissing{0.0} \\
	 & SNB-BI  &  \hylitemissing{0.0} & 61.9 &    4.7 &  52.3 &   14.2 &       80.9 &      \hyliteoverused{100.0} &                     90.4 &                     38.1 &                     80.9 &                     52.3 &    \hylitemissing{0.0} \\
	 & SNB-INT & \hylitemissing{0.0} & 47.3 & \hylitemissing{0.0}& 31.5 & 15.7 & 63.15 & 78.9 & 94.7 & 42.1 & 94.7 & 84.2 &\hylitemissing{0.0} \\
	 \midrule
		  \parbox[t]{2mm} {\multirow{4}{*}{\rotatebox[origin=c]{90}{\textbf{Real}}}}    & FEASIBLE  &                         56.0 &                         58.0 &                        22.0 &                        28.0 &                       40.0 &                        42.0 &                        32.0 &                     58.0 &                     18.0 &                     36.0 &                     16.0 &                   30.0 \\
		                                                                                & Fishmark  &          \hylitemissing{0.0} &          \hylitemissing{0.0} &         \hylitemissing{0.0} &                         9.0 &        \hylitemissing{0.0} &         \hylitemissing{0.0} &         \hylitemissing{0.0} &   \hyliteoverused{100.0} &                     81.8 &                      9.0 &                     72.7 &    \hylitemissing{0.0} \\
		                                                                                & DBPSB     &       \hyliteoverused{100.0} &                         48.0 &                         8.0 &                        32.0 &                       36.0 &         \hylitemissing{0.0} &         \hylitemissing{0.0} &                     68.0 &                     20.0 &                     32.0 &                     20.0 &                   24.0 \\
		                                                                                & BioBench  &                         28.2 &                         25.6 &                        15.3 &                         7.6 &                        7.6 &                        20.5 &                        10.2 &                     71.7 &                     53.8 &                     43.5 &                     38.4 &                   15.3 \\ 
 \bottomrule
	\end{tabular}
\end{table*}

\paragraph{Performance Metrics.} 
This section presents results pertaining to the performance metrics discussed in Section \ref{42-sec:storage-metric}. Table \ref{42-tab:metric-stats} shows the performance metrics used by the selected benchmarks to compare triplestores. The query runtimes for complete benchmark's queries is the central performance metrics and is used by all of the selected benchmarks. In addition, the QpS and QMpH are commonly used in the query processing category. 
We found that in general, the processing overhead generated by query executions is not paid much attention as only SP2Bench measures this metric. In the ``storage'' category, the time taken to load the RDF graph into triplestore is most common. The in-memory/HDD space required to store the dataset and corresponding indexes did not get much attention. The result set correctness and completeness are important metrics to be considered when there is a large number of queries in the benchmark and composite metrics such as QpS and QMpH are used. We can see many of the benchmarks do not explicitly check these two metrics. They mostly assume that the results are complete and correct. However, this might not be the case \cite{largerdfbench2018}. Additionally, only BSBM considers the evaluation of triplestores with simultaneous user requests with updates. However, benchmark execution frameworks such Iguana~\cite{iswc_iguana} can be used to measure the parallel query processing  capabilities of triplestores in presence of multiple querying and update agents. It is further described in Section~\ref{42-sec:iguana}. 

\newcolumntype{P}{>{\centering\arraybackslash}p{0.55cm}}
\begin{table*}[!htb]
\centering
\footnotesize
\caption{Metrics used in the selected benchmarks pertaining to query processing, data storage, result set, simultaneous multiple client requests, and dataset updates. \emph{QpS:} Queries per Second,  \emph{QMpH:} Queries Mix per Hour, \emph{PO:} Processing Overhead, \emph{LT:}	Load Time, \emph{SS:} Storage Space, \emph{IS:} Index Sizes, \emph{RCm:} Result Set Completeness, \emph{RCr:} Result Set Correctness,  \emph{MC:}	Multiple Clients,  \emph{DU:} Dataset Updates.}
\label{42-tab:metric-stats}
\begin{tabular}{@{}llPPPPPPPPPP@{}}
	\toprule && \multicolumn{3}{c}{Processing} & \multicolumn{3}{c}{Storage} & \multicolumn{2}{c}{Result Set} & \multicolumn{2}{c}{Additional}  \\ 
	\cmidrule(lr){3-5} \cmidrule(lr){6-8} \cmidrule(lr){9-10} \cmidrule(lr){11-12} 
	&Benchmark &	QpS	 &	QMpH &	PO &		LT &		SS  &		IS	 &	RCm	 &	RCr &		MC &		DU \\
	\midrule
	\parbox[t]{2.7mm} {\multirow{6}{*}{\rotatebox[origin=c]{90}{\textbf{Synthetic}}}}&Bowlogna  &  	\xmark & 	\xmark & 	\xmark & 		\cmark & 	\xmark & 	\cmark & 	\xmark & 	\xmark & 	\xmark & 	\xmark \\
& TrainBench &   	\xmark & 	\xmark & 	\xmark & 		\cmark & 	\xmark & 	\xmark & 	\cmark & 	\cmark & 	\xmark & 	\cmark\\
& BSBM	   &    \cmark & 	\cmark & 	\xmark & 		\cmark & 	\xmark & 	\xmark & 	\cmark & 	\cmark & 	\cmark & 	\cmark\\
& SP2Bench  &    	\xmark & 	\xmark & 	\cmark & 		\cmark & 	\cmark & 	\xmark & 	\cmark & 	\cmark & 	\xmark & 	\xmark\\
& WatDiv	   &    \xmark & 	\xmark & 	\xmark & 		\xmark & 	\xmark & 	\xmark & 	\xmark & 	\xmark & 	\xmark & 	\xmark\\
& SNB-BI &  	\cmark & 	\cmark & 	\xmark & 		\xmark & 	\xmark & 	\xmark & 	\cmark & 	\cmark & 	\xmark & 	\xmark\\
& SNB-INT & 	\cmark & 	\cmark & 	\xmark & 		\xmark & 	\xmark & 	\xmark & 	\cmark & 	\cmark & 	\xmark & 	\cmark\\ \midrule
 \parbox[t]{2mm} {\multirow{4}{*}{\rotatebox[origin=c]{90}{\textbf{Real}}}}    & FEASIBLE & 	    \cmark & 	\cmark & 	\xmark & 		\xmark & 	\xmark & 	\xmark & 	\cmark & 	\cmark & 	\xmark & 	\xmark\\
& Fishmark & 	    \cmark & 	\xmark & 	\xmark & 		\xmark & 	\xmark & 	\xmark & 	\xmark & 	\xmark & 	\xmark & 	\xmark\\
& DBPSB &         \cmark & 	\cmark & 	\xmark & 		\xmark & 	\xmark & 	\xmark & 	\xmark & 	\xmark & 	\xmark & 	\xmark\\
& BioBench & 	    \xmark & 	\xmark & 	\xmark & 		\cmark & 	\cmark & 	\xmark & 	\cmark & 	\xmark & 	\cmark & 	\xmark\\
 \bottomrule
\end{tabular}
\end{table*}

\subsubsection{Summary of Storage Benchmarks}
We performed a comprehensive analysis of existing benchmarks by studying synthetic and real-world datasets as well as by employing SPARQL queries with multiple variations. Our evaluation results suggest the following:
\begin{enumerate}
\item The dataset structuredness problem is well covered in recent synthetic data generators (e.g., WatDiv, TrainBench). The low relationship specialty problem in synthetic datasets still exists in general and needs to be covered in future synthetic benchmark generation approaches.
\item The FEASIBLE framework employed on DBpedia generated the most diverse benchmark in our evaluation.
\item The SPARQL query features we selected have a weak correlation with query execution time, suggesting that the query runtime is a complex measure affected by multi-dimensional SPARQL query features. Still, the number of projection variables, join vertices, triple patterns, the result sizes, and the join vertex degree are the top five SPARQL features that most impact the overall query execution time.
\item Synthetic benchmarks often fail to contain important SPARQL clauses such as \texttt{DISTINCT}, \texttt{FILTER}, \texttt{OPTIONAL}, \texttt{LIMIT} and \texttt{UNION}.
\item The dataset structuredness has a direct correlation with the result sizes and execution times of queries and indirect correlation with dataset specialty.
\end{enumerate}

\subsection{Manual Revision \& Authoring}

Using SPARQL queries, the user can interact with the stored data directly, e.g., by correcting errors or adding additional information~\cite{Ngomo2014}. This is the third step of the Linked Data Lifecycle. However, since it mainly comprises user interaction, there are no automatic benchmarks for tools supporting this step. Hence, we will not look further into it.

\subsection{Linking }
\label{42-sec:linking}

With cleaned data originating from different sources, the generation of links between the different information assets need to be established~\cite{Ngomo2014}. The generation of such links between knowledge bases is one of the key steps of the Linked Data publication process.\footnote{\url{http://www.w3.org/DesignIssues/LinkedData.html}}  A plethora of approaches has thus been devised to support this process~\cite{DBLP:journals/semweb/NentwigHNR17}.

The formal specification of Link Discovery adopted herein is akin to that proposed in \cite{DBLP:journals/jodsn/Ngomo12}.
Given two (not necessarily distinct) sets $\mathcal{S}$ resp. $\mathcal{T}$ of source resp. target resources as well as a relation $R$,
the goal of LD is is to find the set $M = \{(s, \tau) \in \mathcal{S} \times \mathcal{T}: R(s, \tau)\}$ of pairs $(s, \tau) \in \mathcal{S} \times \mathcal{T}$ such that $R(s, \tau)$.
In most cases, computing $M$ is a non-trivial task.
Hence, a large number of frameworks (e.g., SILK~\cite{DBLP:conf/webdb/IseleJB11}, LIMES~\cite{DBLP:journals/jodsn/Ngomo12} and KnoFuss~\cite{DBLP:conf/kcap/NikolovUM07}) aim to approximate $M$ by computing the \emph{mapping} $M' = \{(s,\tau) \in \mathcal{S} \times \mathcal{T}: \sigma(s, \tau) \geq \theta\}$, where $\sigma$ is a similarity function and $\theta$ is a similarity threshold. For example, one can configure these frameworks to compare the dates of birth, family names and given names of persons across census records to determine whether they are duplicates.
We call the equation which specifies $M'$ a \emph{link specification} (short LS; also called linkage rule in the literature, see e.g., \cite{DBLP:conf/webdb/IseleJB11}).
Note that the Link Discovery problem can be expressed equivalently using distances instead of similarities in the following manner: Given two sets $\mathcal{S}$ and $\mathcal{T}$ of instances, a (complex) distance measure $\delta$ and a distance threshold $\theta \in [0, \infty[$, determine $M' = \{(s,\tau) \in \mathcal{S} \times \mathcal{T}: \delta(s, \tau) \leq \vartheta\}$.\footnote{Note that a distance function $\delta$ can always be transformed into a normed similarity function $\sigma$ by setting $\sigma(x, y) = (1+\delta(x, y))^{-1}$. Hence, the distance threshold $\vartheta$ can be transformed into a similarity threshold $\theta$ by means of the equation $\theta = (1+\vartheta)^{-1}$.}

Under this so-called \emph{declarative paradigm}, two entities $s$ and $\tau$ are then considered to be linked via $R$ if $\sigma(s, \tau) \geq \theta$. Na\"ive algorithms require $O(|\mathcal{S}||\mathcal{T}|) \in O(n^2)$ computations to output $M'$. 
Given the large size of existing knowledge bases, time-efficient approaches able to reduce this runtime are hence a central component of efficient link discovery, as link specifications need to be computed in acceptable times. This efficient computation is in turn the proxy necessary for machine learning techniques to be used to optimize the choice of appropriate $\sigma$ and $\theta$ and thus ensure that $M'$ approximates $M$ well even when $M$ is large~\cite{DBLP:journals/semweb/NentwigHNR17}). 

Benchmarks for link discovery frameworks use tasks of the form $t_i = ((\mathcal{S}_i,\mathcal{T}_i,R_i),M_i)$. Depending on the given datasets $\mathcal{S}_i$ and $\mathcal{T}_i$, and the given relation $R_i$, a benchmark is able to make different demands on the Link discovery frameworks. To this end, such frameworks are designed to accommodate a large number of link discovery approaches within a single extensible architecture to address two main challenges measured by the benchmark:
\begin{enumerate}
	\item \emph{Time-efficiency}: The mere size of existing knowledge bases (e.g., $30+$ billion triples in LinkedGeoData~\cite{DBLP:journals/semweb/StadlerLHA12}, $20+$ billion 		triples in LinkedTCGA~\cite{DBLP:conf/i-semantics/SaleemPNADD13}) makes efficient solutions indispensable to the use of link discovery frameworks in real application scenarios. LIMES for example addresses this challenge by providing time-efficient approaches based on the characteristics of metric spaces~\cite{DBLP:conf/ijcai/NgomoA11,DBLP:conf/semweb/Ngomo12}, orthodromic spaces~\cite{DBLP:conf/semweb/Ngomo13} and filter-based paradigms~\cite{DBLP:conf/semweb/SoruN13}.
	\item \emph{Accuracy}: The second key performance indicator is the accuracy provided by link discorvery frameworks. Efficient solutions are of little help if the results they generate are inaccurate.
	Hence, LIMES also accommodates dedicated machine-learning solutions that allow the generation of links between knowledge bases with a high accuracy.
	These solutions abide by paradigms such as batch and active learning~\cite{DBLP:conf/semweb/NgomoLAH11,DBLP:conf/esws/NgomoL12,DBLP:conf/semweb/NgomoL13}, unsupervised learning~\cite{DBLP:conf/semweb/NgomoL13} and even positive-only learning~\cite{WOMBAT_2017}.
\end{enumerate}

\emph{The Ontology Evaluation Alignment Initiative (OAEI)}\footnote{\url{http://www.ontologymatching.org}} has performed yearly contests
since 2005 to comparatively evaluate current tools for ontology and instance
matching. 
The original focus has been on ontology matching, but since 2009 instance matching has also been a regular evaluation track. 
Table \ref{42-tab:oaei} (From~\cite{DBLP:journals/semweb/NentwigHNR17}) gives an overview of the OAEI instance matching tasks in five contests from 2010 to 2014.
Most tasks have only been used in one year while others like IIMB have been changed in different years.
The majority of the linking tasks are based on synthetic datasets where values and the structural context of instances have been modified in a controlled way. 
Linking tasks cover different domains (life sciences, people, geography, etc.) and data sources (\emph{DBpedia, Freebase, GeoNames, NYTimes,} etc.). 
Frequently, the benchmarks consist of several interlinking tasks to cover a certain spectrum of complexity. 
The number of instances are rather small in all tests with $9,958$ the maximum size of data source. 
The evaluation focus has been solely on the effectiveness (e.g., F-Measure) while runtime efficiency has not been measured.
Almost all tasks focus on identifying equivalent instances (i.e., \texttt{owl:sameAs} links).  

We briefly characterize the different OAEI tasks as follows:
\begin{itemize}
\item \emph{IIMB and Sandbox (SB).} The IIMB benchmark has been part of the 2010,
2011 and 2012 contests and consists of 80 tasks using synthetically modified datasets derived from instances of 29 Freebase concepts.  
The number of instances varies from year to year but the instances have a very small size (e.g., at most 375 instances in 2012).  
The  Sandbox (SB) benchmark from 2012 is very similar to IIMB but limited to 10 different tasks~\cite{Aguirre2012}.

\item \emph{PR (Persons/Restaurant).} This benchmark is based on real-person and restaurant-instance data that are artificially modified by adding duplicates and variations of property values. 
With 500--600 instances for the restaurant data and
even less in the person data source, the dataset is relatively small.~\cite{Euzenat2010}

\item \emph{DI-NYT (Data Interlinking - NYT).} This 2011 benchmark includes seven tasks to link about $10,000$ instances from the NYT data source to DBpedia, Freebase and
GeoNames instances.
The perfect match result contains about 31,000
\texttt{owl:sameAs} links to be identified~\cite{Euzenat2011a}.

\item \emph{RDFT.} This 2013 benchmark is also of small size (430 instances) and uses several tests with differently modified DBpedia data. 
For the first time in the OAEI instance matching track, no reference mapping is provided for the actual evaluation task.
Instead, training data with an appropriate reference mapping is given for each test case thereby supporting frameworks relying on supervised learning~\cite{Grau2013}.

\item \emph{OAEI 2014.} Two benchmark tasks have to be performed in 2014. 
The first one (id-rec) requiring the identification of the  same real-world book entities (\texttt{owl:sameAs} links). 
For this purpose, 1,330 book instances have to be matched with with 2,649 synthetically modified instances in the target dataset.
Data transformations include changes like the substitution of book titles and labels with keywords as well as language transformations.
In the second task (sim-rec) similarity of pairs of instances are determined, which do not reflect the same real-world entities. 
This addresses common preprocessing tasks, e.g., to reduce the search space for \gls{ld}. In 2014, the evaluation platform SEALS~\cite{Garcia-Castro2011}, which has been used for ontology matching in previous years, is used for instance matching, too (see Section~\ref{42-sec:seals}).

\end{itemize}

\begin{table*}
	\centering
	\caption{OAEI instance matching tasks over the years. ``--'' means not existing, ``?'' unclear from publication~\cite{DBLP:journals/semweb/NentwigHNR17}.}
	\label{42-tab:oaei}
	\begin{tabularx}{\textwidth}{@{}lllllllll@{}}
		\toprule
		\parbox[t]{2mm}{\multirow{2}{*}{\rotatebox[origin=c]{90}{Year}}} & \multirow{2}{*}{Name}    & Input  & Problem    & \multirow{2}{*}{Domains}       & \multirow{2}{*}{LOD Sources} & \multirow{2}{*}{Link Type}    & Max. \# & \multirow{2}{*}{Tasks} \\
		&         & Format &  Type   &               & &              & Resources  & \\
		\toprule
		\parbox[t]{2mm}{\multirow{7}{*}{\rotatebox[origin=c]{90}{2010}}}            & DI      & RDF    & real       & life sciences & diseasome   & equality     & 5,000   & 4 \\ 
		&         &        &            &               & drugbank    &              & \\
		&         &        &            &               & dailymed    &              & \\
		&         &        &            &               & sider       &              & \\
		\cmidrule(r){2-9}
		& IIMB    & OWL    & artificial & cross-domain  & Freebase    & equality                 & 1,416   & 80 \\ 

		\cmidrule(r){2-9}
		& PR      & RDF,   & artificial & people        & --           & equality                  & 864     & 3 \\ 
		&         & OWL    &            & geography     &             &                         & \\
		\midrule
		\parbox[t]{2mm}{\multirow{5}{*}{\rotatebox[origin=c]{90}{2011}}}              & DI-NYT  & RDF    & real       & people        & NYTimes     & equality                 & 9,958   & 7 \\
		&         &        &            & geography     & DBpedia     &                      & \\
		&         &        &            & organizations & Freebase    &              & \\
		&         &        &            &               & Geonames    &              & \\
		\cmidrule(r){2-9}
		& IIMB    & OWL    & artificial & cross-domain  & Freebase    & equality                  & 1,500   & 80\\
		\midrule
		\parbox[t]{2mm}{\multirow{2}{*}{\rotatebox[origin=c]{90}{2012}}}             & SB      & OWL    & artificial & cross-domain  & Freebase    & equality                  & 375     & 10\\ 
		\cmidrule(r){2-9}
		& IIMB    & OWL    & artificial & cross-domain  & Freebase    & equality                  & 375     & 80\\
		\midrule
		\parbox[t]{2mm}{\rotatebox[origin=c]{90}{2013}}              & RDFT    & RDF    & artificial & people        & DBpedia     & equality                  & 430     & 5\\
		\midrule
		\parbox[t]{2mm}{\multirow{2}{*}{\rotatebox[origin=c]{90}{2014}}}               & id-rec  & OWL    & artificial & publications  & ?           & equality                 & 2,649   & 1\\
		\cmidrule(r){2-9}
		& sim-rec & OWL    & artificial & publications  & ?           & similarity        & 173     & 1 \\
		\bottomrule
	\end{tabularx}
\end{table*}

\subsection{Enrichment}


Linked Data Enrichment is an important topic for all applications that rely on a large number of knowledge bases and necessitate a unified view on this data, e.g., Question Answering frameworks, Linked Education and all forms of semantic mashups.
In recent work, several challenges and requirements to Linked Data consumption and integration have been pointed out~\cite{MIL+10}.
Several approaches and frameworks have been developed with the aim of addressing many of these challenges. 
For example, the R2R framework~\cite{bizer2010r2r} addresses those by allowing mappings to be published across knowledge bases, mapping classes defining and property value transformation.
While this framework supports a large number of transformations, it does not allow the automatic discovery of possible transformations.
The Linked Data Integration Framework LDIF~\cite{SCH+11}, whose goal is to support the integration of RDF data, builds upon R2R mappings and technologies such as SILK~\cite{DBLP:conf/webdb/IseleJB11} and LDSpider\footnote{\url{http://code.google.com/p/ldspider/}}. 
The concept behind the framework is to enable users to create periodic integration jobs via simple XML configurations. 
These configurations however have to be created manually.
The same drawback holds for the Semantic Web Pipes\footnote{\url{http://pipes.deri.org/}}, which follows the idea of Yahoo Pipes\footnote{\url{http://pipes.yahoo.com/pipes/}} to enable the integration of data in formats such as RDF and XML. 
By using Semantic Web Pipes, users can efficiently create semantic mashups by using a number of operators (such as getRDF, getXML, etc.) and connecting these manually within a simple interface.
It begins by detecting URIs that stand for the same real-world entity and either merging them to one or linking them via \texttt{owl:sameAs}.
Fluid Operations' Information Workbench\footnote{\url{http://www.fluidops.com/information-workbench/}} allows users to search through, manipulate and integrate for purposes such as business intelligence.
The work presented in~\cite{choudhury2009enrichment} describes a framework for semantic enrichment, ranking and integration of web videos.
\cite{abel2011semantic} presents semantic enrichment framework of \emph{Twitter} posts.
 \cite{hasan2011toward} tackles the linked data enrichment problem for sensor data 
via an approach that sees enrichment as a process driven by situations of interest. 

To the best of our knowledge, \deer~\cite{DEER_2015} is the first generic approach tailored towards learning enrichment pipelines of Linked Data given a set of atomic enrichment functions.
\deer models RDF dataset enrichment workflows as ordered sequences of enrichment functions. 
These enrichment functions take as input exactly one RDF dataset and return an altered version of it by virtue of addition and deletion of triples, as any enrichment operation on a given dataset can be represented as a set of additions and deletions.

The benchmarking proposed by~\cite{DEER_2015}  aims to quantify how well the presented RDF data enrichment approach can automate the enrichment process.
The authors  thus assumed being given manually created training examples and having to reconstruct a possible enrichment pipeline to generate target concise bounded descriptions (CBDs\footnote{\url{https://www.w3.org/Submission/CBD/}}) resources from the source CBDs.
Therefore, they used three publicly available datasets for generating the benchmark datastes:
\begin{enumerate}
	\item  From the biomedical domain, the authors chose \emph{DrugBank}\footnote{\emph{DrugBank} is the Linked Data version of the DrugBank database, which is a repository of almost 5000 FDA-approved small molecule and biotech drugs, for RDF dump see  \url{http://wifo5-03.informatik.uni-mannheim.de/drugbank/drugbank_dump.nt.bz2}}.
	They chose this dataset because it is linked with many other datasets\footnote{See \url{http://datahub.io/dataset/fu-berlin-drugbank} for complete list of linked dataset with \emph{DrugBank}.}
	from which the authors extracted enrichment data using their atomic enrichment functions.
	For evaluation, they deployed a manual enrichment pipeline $M_{manual}$, where they enriched the drug data found in \emph{DrugBank} using abstracts dereferenced from \emph{DBpedia}.
	Then, they conformed both \emph{DrugBank} and \emph{DBpedia} source authority URIs to one unified URI.
	For \emph{DrugBank} they manually deployed two experimental pipelines:
	\begin{itemize}
		\item $M_{DrugBank}^1 = (m_{1}, m_{2})$, where  $m_{1}$ is  a dereferencing function that dereferences any \texttt{dbpedia-owl:ab\-stract} from DBpedia and $m_{2}$ is an authority conformation function that conforms the \emph{DBpedia} subject authority\footnote{\url{http://dbpedia.org}} to the target subject authority of \emph{DrugBank}\footnote{\url{http://wifo5-04.informatik.uni-mannheim.de/drugbank/resource/drugs}}.
		\item $M_{DrugBank}^2 = M_{DrugBank}^1 \doubleplus m_{3} $, where $m_{3}$ is an authority conformation function that conforms \emph{DrugBank}'s authority to the \emph{Example} authority\footnote{\url{http://example.org}}.
	\end{itemize}
	\item From the music domain, the authors chose the \emph{Jamendo}\footnote{\emph{Jamendo} contains a large collection of music-related information about artists and recordings, for RDF dump see \url{http://moustaki.org/resources/jamendo-rdf.tar.gz}} dataset. 
	They selected this dataset as it contains a substantial amount of embedded information hidden in literal properties such as \texttt{mo:biography}.
	The goal of their enrichment process is to add a geospatial dimension to \emph{Jamendo}, e.g., location of a recording or place of birth of a musician.
	To this end, the authors of~\cite{DEER_2015} deployed a manual enrichment pipeline, in which they enriched \emph{Jamendo}'s music data by adding additional geospatial data found by applying the NER of the NLP enrichment function against \texttt{mo:biography}.
	For \emph{Jamendo} they manually deployed one experimental pipeline:
	\begin{itemize}
		\item $M_{Jamendo}^1 = \{m_{4}\}$, where  $m_{4}$ is an enrichment function that finds \emph{locations} in \texttt{mo:biography} using natural language processing (NLP) techniques.
	\end{itemize}

	\item From the multi-domain knowledge base \emph{DBpedia}~\cite{dbpedia-swj} the authors selected the class \texttt{Ad\-ministrativeRegion} to be our third dataset. 
	As DBpedia is a knowledge base with a complex ontology, they build a set of 5 pipelines of increasing complexity:
	\begin{itemize}
		\item $M_{DBpedia}^1 = \{m_5\}$, where  $m_5$ is an authority conformation function that conforms the \emph{DBpedia} subject authority to the \emph{Example} target subject authority.
		\item $M_{DBpedia}^2 = m_6 \doubleplus M_{DBpedia}^1 $, where $m_6$ is a dereferencing function that dereference any \texttt{dbpedia-owl:ideo\-lo\-gy} and $\doubleplus$ is the list append operator.
		\item $M_{DBpedia}^3 = M_{DBpedia}^2 \doubleplus m_7 $, where $m_7$ is a NLP function that finds \emph{all} named entities in \texttt{dbpedia-owl:abstract}.
		\item $M_{DBpedia}^4 = M_{DBpedia}^3 \doubleplus m_8 $, where $m_8$ is a filter function that filters for abstracts.
		\item $M_{DBpedia}^5 = M_{DBpedia}^3 \doubleplus m_9 $, where $m_9$ is a predicate conformation function that conforms the source predicate \texttt{dbpedia-owl:abstract} to the target predicate of \texttt{dcterms:abstract}.
	\end{itemize}
\end{enumerate}
\label{sec:pipeline}

Altogether, The authors manually generated a set of 8 pipelines, which they then applied against their respective datasets. The resulting pairs of CBDs were used to generate the positive examples, which were simply randomly selected pairs of distinct CBDs. 
All generated pipelines are available at the project web site.\footnote{\url{https://github.com/GeoKnow/DEER/evaluations/pipeline_learner}}
\subsection{Quality Analysis}

Since Linked Data origins from different publishers, a variety of information quality exists and different strategies exists to assess them~\cite{Ngomo2014}. Since several previous steps are extracting and generating triples automatically, an important part of the quality analysis is the evaluation of single facts with respect to their veracity. These fact validation systems (also known as fact checking systems) take a triple $(s, p, o)$ as input $i_j$ and return a veracity score as output~\cite{syed2019copaal}.

An extension to the fact validation task is the temporal fact validation. A temporal fact validation system takes a point in time as additional input. It validates whether the given fact was true at the given point in time. However, to the best of our knowledge,~\cite{Gerber2015DeFacto} is the only approach that tackles this task and FactBench is the only dataset available for benchmarking such a system.

\subsubsection{Validation system types}

Most of the fact validation systems use reference knowledge to validate a given triple. There are two major types of knowledge which can be used---structured and unstructured knowledge~\cite{syed2019copaal}. 

Approaches relying on structured knowledge use a reference knowledge base to search for patterns or paths supporting or refuting the given fact. Other approaches transfer the knowledge base into an embedding space and calculate the similarity between the entities positions in the space and the positions the entities should have if the fact is true~\cite{syed2019copaal}. For all these approaches, it is necessary that the single IRIs of the fact's subject, predicate and object are available in the knowledge base. Hence, the knowledge base the facts of a fact validation benchmark rely on is background knowledge a benchmarked fact validation system needs to have.

The second group of approaches is based on unstructured---in most cases textual---knowledge~\cite{Gerber2015DeFacto,syed2019copaal}. These approaches are generating search queries based on the give fact and search for documents providing evidences for the fact. To this end, it is necessary for such an approach to be aware of the subject's and object's label. Additionally, a pattern how to formulate a fact with respect to the given predicate might be necessary.

\subsubsection{Datasets}

Several datasets have been created during the last years. Table~\ref{42-tab:FCDatasets} gives an overview. \emph{FactBench}~\cite{Gerber2015DeFacto} is a collection of 7 datasets.\footnote{\url{https://github.com/DeFacto/FactBench}} The facts of the datasets have a time range attached which shows at which point in time the facts are true. The datasets share 1500 true facts selected from DBpedia and Freebase using 10 different properties.\footnote{The dataset has 150 true and false facts for each of the following properties: \emph{award}, \emph{birthPlace}, \emph{deathPlace}, \emph{foundationPlace}, \emph{leader}, \emph{team}, \emph{author}, \emph{spouse}, \emph{starring} and \emph{subsidiary}.} However, the datasets differ in the way the 1500 false facts are derived from true facts.
\begin{itemize}
\item \emph{Domain}: The subject is replaced by a random resource.
\item \emph{Range}: The object is replaced by a random resource.
\item \emph{Domain-Range}: Subject and object are replaced as described above.
\item \emph{Property}: The predicate of the triple is replaced with a randomly chosen property.
\item \emph{Random}: A triple is generated randomly.
\item \emph{Date}: The time range of a fact is changed. If the range is a single point in time, it is randomly chosen from a gaussian distribution which has its mean at the original point in time and a variance of 5 years. For time intervals, the start year and the duration are drawn from two gaussion distributions.
\item \emph{Mix}: A mixture comprising one sixth of each of the sets above.
\end{itemize}
It has to be noted that all changes are done in a way that makes sure that the subject and object of the generated false triple follow the domain and range restrictions of the fact's predicate. In all cases, the generated fact must not be made present in the knowledge base.

In~\cite{Shi2016PredPath}, the authors propose several datasets based on triples from the DBpedia. \emph{CapitalOf \#1} comprises \emph{capitalOf} relations of the 5 largest US cities cities of the 50 US states. If the correct city is not within the top 5 cities, an additional fact with the correct city is added. 
\emph{CapitalOf \#2} comprises the same 50 correct triples. Additionally, for each capital four triples with a wrong state are created. 
The \emph{US Civil War} dataset comprises triples connecting military commanders to decisive battles of the US Civil war. 
The \emph{Company CEO} dataset comprises 205 correct connections between CEOs and their company using the \emph{keyPerson} property. 1025 incorrect triples are created by randomly selecting a CEO and a company from the correct facts.
The \emph{NYT Bestseller} comprises 465 incorrect random pairs of authors and books as well as 93 true triples of the New York Times bestseller lists between 2010 and 2015.
The \emph{US Vice-President} contains 47 correct and 227 incorrect pairs of vice-presidents and presidents of the United States.
Additionally, the authors create two datasets based on SemMedDB. The \emph{Disease} dataset contains triples marking an amino acid, peptide or protein as a cause of a disease or syndrome. It comprises 100 correct and 457 incorrect triples.
The \emph{Cell} dataset connects cell functions with the gene that causes them. It contains 99 correct and 435 incorrect statements.

In~\cite{shiralkar2017finding}, the authors propose several datasets. \emph{FLOTUS} comprises US presidents and their spouses. The\emph{NBA-Team} contains triples assigning NBA players to teams. The \emph{Oscars} dataset comprises triples related to movies.
Additionally, the authors reuse triples from the Google Relation Extraction Corpora (\emph{GREC}) containing triples about the \emph{birthplace}, \emph{deathplace}, \emph{education} and \emph{institution} of famous people. The ground truth for these triples is derived using crowd sourcing.
The \emph{profession} and \emph{nationality} datasets are derived from the WSDM Cup 2017 Triple Scoring challenge. Since the challenge only provides true facts, false facts have been randomly drawn.

In~\cite{Syed2018FactCheck}, the authors claim that several approaches for fact validation do not take the predicate into account and that for most of the existing benchmarking datasets, checking the connection between the subject and object is already sufficient. For creating a more difficult dataset, the authors query a list of persons from DBpedia that have their birth- and deathplaces in two different countries. They use these triples as true facts and swap birth- and deathplace to create wrong facts.

\begin{table*}
\centering
\caption{Summary of benchmark datasets for fact validations.}
\label{42-tab:FCDatasets}
\begin{tabular}{@{}lrrrc@{}}
\toprule
\multirow{2}{*}{\textbf{Dataset}} & 
\multicolumn{3}{c}{\textbf{Facts}} & \multirow{2}{*}{\textbf{Temporal}} \\
\cmidrule{2-4}
& \multicolumn{1}{c}{True} & \multicolumn{1}{c}{False} & \multicolumn{1}{c}{Total} & \\
\midrule
Birthplace/Deathplace (\cite{Syed2018FactCheck}) & 206 & 206 & 412 & no \\
CapitalOf \#1 (\cite{Shi2016PredPath}) & 50 & 209 & 259 & no \\
CapitalOf \#2 (\cite{Shi2016PredPath}) & 50 & 200 & 250 & no \\
Cell (\cite{Shi2016PredPath}) & 99 & 435 & 534 & no \\
Company CEO (\cite{Shi2016PredPath}) & 205 & 1025 & 1230 & no \\
Disease (\cite{Shi2016PredPath}) & 100 & 457 & 557 & no \\
FactBench Date (\cite{Gerber2015DeFacto}) & 1\,500 & 1\,500 & 3\,000 & yes \\
FactBench Domain (\cite{Gerber2015DeFacto}) & 1\,500 & 1\,500 & 3\,000 & yes \\
FactBench Domain-Range (\cite{Gerber2015DeFacto}) & 1\,500 & 1\,500 & 3\,000 & yes \\
FactBench Mix (\cite{Gerber2015DeFacto}) & 1\,500 & 1\,560 & 3\,060 & yes \\
FactBench Property (\cite{Gerber2015DeFacto}) & 1\,500 & 1\,500 & 3\,000 & yes \\
FactBench Random (\cite{Gerber2015DeFacto}) & 1\,500 & 1\,500 & 3\,000 & yes \\
FactBench Range (\cite{Gerber2015DeFacto}) & 1\,500 & 1\,500 & 3\,000 & yes \\
FLOTUS (\cite{shiralkar2017finding}) & 16 & 240 & 256 & no \\
GREC-Birthplace (\cite{shiralkar2017finding}) & 273 & 819 & 1\,092 & no \\
GREC-Deathplace (\cite{shiralkar2017finding}) & 126 & 378 & 504 & no \\
GREC-Education (\cite{shiralkar2017finding}) & 466 & 1\,395 & 1\,861 & no \\
GREC-Institution (\cite{shiralkar2017finding}) & 1\,546 & 4\,638 & 6\,184 & no \\
NBA-Team (\cite{shiralkar2017finding}) & 41 & 143 & 164 & no \\
NYT Bestseller (\cite{Shi2016PredPath}) & 93 & 465 & 558 & no \\
Oscars (\cite{shiralkar2017finding}) & 78 & 4\,602 & 4\,680 & no \\
US Civil War (\cite{Shi2016PredPath}) & 126 & 584 & 710 & no \\
US Vice-President (\cite{Shi2016PredPath}) & 47 & 227 & 274 & no \\
WSDM-Nationality (\cite{shiralkar2017finding}) & 50 & 150 & 200 & no \\
WSDM-Profession (\cite{shiralkar2017finding}) & 110 & 330 & 440 & no \\
\bottomrule
\end{tabular}
\end{table*}

\subsubsection{Key Performance Indicators}

The performance of a fact validation system can be measured regarding its effectiveness and efficiency. To measure the effectiveness, the Area under ROC is a common choice~\cite{syed2019copaal}. Interpreting the validation as a binary classification task allows the usage of Precision, Recall and F1-measure. However, this comes with the search of a threshold for the minimal veracity value that can still be accepted as true~\cite{Syed2018FactCheck}. The efficiency is typically measured with respect to the runtime of a fact validation system~\cite{syed2019copaal}.

\subsection{Evolution \& Repair}

If problems within a \gls{kg} are detected, repairing and managing the evolution of the data is necessary~\cite{Ngomo2014}. Since repair tools mainly support a manual task, and are therefore hard to benchmark automatically, we will focus on the benchmarking of versioning systems. These systems work similarly to SPARQL stores described in Section~\ref{42-sec:storage}. However, instead of answering a query based on a single RDF graph, they allow the querying of data from a single or over different versions of the graph~\cite{Fernandez2019Bear}.

With respect to the proposed benchmark schema, the benchmarks for versioning systems are very similar to the triple store benchmarks. A benchmark dataset comprises several versions of a knowledge base, a set of queries and a set of expected results for these queries. The knowledge base versions can be seen as background data that is provided at the beginning of the benchmarking. The input for a task is a single query and the expected output is the expected query result. The benchmarked versioning system has to answer the query based on the provided knowledge base versions.

\subsubsection{Query Types}

In the literature, different types of queries are defined, which a versioning system needs to be able to handle. The authors in \cite{Fernandez2019Bear} provide the following 6 queries:
\begin{itemize}
\item \emph{Version materialisation queries} retrieve the full version of a knowledge base for a given point in time.
\item \emph{Single-version structured queries} retrieve a subset of the data of a single version.
\item \emph{Cross-version structured queries} retrieve data across several versions.
\item \emph{Delta materialisation queries} retrieve the differences between two or more versions.
\item \emph{Single-delta structured queries} retrieve data from a single delta of two versions.
\item \emph{Cross-delta structured queries} retrieve data across several deltas.
\end{itemize}
The authors in~\cite{Papakonstantinou2016Versioning,Papakonstantinou2017SPBvBL} create a more fine-grained structure providing 8 different types of queries.

\subsubsection{Datasets}

The BEAR benchmark dataset proposed in \cite{Fernandez2019Bear} comprises three parts---BEAR-A, BEAR-B, and BEAR-C. BEAR-A comprises 58 versions of a single real-world knowledge base taken from the Dynamic Linked Data Observatory.\footnote{http://swse.deri.org/dyldo/} The set of queries comprises triple pattern queries and is manually created by the authors to cover all basic functionalities a versioning system has to provide. Additionally, it is ensured that the queries give different results for all versions while they are $\epsilon$-stable, i.e., the cardinalities of the results across the versions never exceed $(1 \pm \epsilon)$ of the mean cardinality.

The BEAR-B dataset relies on the DBpedia Live endpoint. The authors of~\cite{Fernandez2019Bear} took the changesets from August to October 2015 reflecting changes in the DBpedia triggered by edits in the Wikipedia. The dataset comprises data of the 100 most volatile resources and the changes to it. The data is provided in three granularities---the \emph{instant} application of changes leads to 21,046 different versions. Additionally, the authors applied summarizations \emph{hourly} and \emph{daily}, leading to 1,299 and 89 versions, respectively. The queries for this dataset are taken from the LSQ dataset~\cite{lsq2015}.

The BEAR-C dataset relies on 32 snapshots of the European Open Data portal.\footnote{http://data.europa.eu/euodp/en/data/} Each snapshot comprises roughly 500m triples describing datasets of the open data portal. The authors create 10 queries that should be too complex to be solved by existing versioning systems ``in a straightforward and optimized way''~\cite{Fernandez2019Bear}. 

Evogen is a dataset generator proposed in~\cite{Meimaris2016}. It relies on the LUBM generator~\cite{lubm2005} and generates additional versions based on its configuration. Based on which delta the generator has used to create different versions, the 14 LUBM queries are adapted to the generated data.

The SPBv dataset generator proposed in~\cite{Papakonstantinou2017SPBvBL} is an extension of the SPB generator~\cite{Kotsev2016}, creating RDF based on the BBC core ontology. It mimics the evolution of journalistic articles in the world of online journalism. While new articles are published, already existing articles are changed over time and different versions of articles are created.

\subsubsection{Key Performance Indicators}

The main key performance indicator used for versioning systems is the query runtime. Analysing these runtimes shows for which query type a certain versioning system has a good or bad performance. A second indicator is the space used by the versioning system to store the data~\cite{Fernandez2019Bear,Papakonstantinou2017SPBvBL}. In addition to,~\cite{Papakonstantinou2017SPBvBL} proposes the measurement of the time a system needs to store the initial version of the knowledge base and the time it needs to apply the changes to create newer versions.


\subsection{Search, Browsing \& Exploration}

The user has to be enabled to access the \gls{ld} in a fast and user friendly way~\cite{Ngomo2014}. This need has led to the development of keyword search tools and question answering systems for the Web of Data. As a result of the interest in these systems, a large number of evaluation campaigns have been undertaken~\cite{usbeck2019gerbilQA}. For example, the TREC conference started a question answering track to provide domain-independent evaluations over unstructured corpora~\cite{voorhees1999trec}. Another series of challenges is the BioASQ series~\cite{bioasq}, which seeks to evaluate QA systems on biomedical data. In contrast to previous series, systems must use RDF data in addition to textual data. For the NLQ shared task, a dataset has been released that is answerable purely by DBpedia and SPARQL.
Another series of challenges is the Question Answering over Linked Data (QALD) campaign~\cite{qald4,qald5,qald6}. It includes different types of question answering benchmarks that are (1) purely based on RDF data, (2) based on RDF and textual data, (3) statistical data, (4) data from multiple \glspl{kb} or (5) based in the music-domain.

Alongside the main question answering task (QA), the authors of~\cite{usbeck2019gerbilQA} define 5 sub tasks that are used for deep analysis of the performance of question answering systems.
\begin{itemize}
\item \emph{QA}: The classical question answering task uses a plain question $q_i$ as input and expects a set of answers $A_i$ as output. Since the answering of questions presumes certain knowledge, most question answering datasets define a knowledge base $K$ as background knowledge for the task. It should be noted that $A_i$ might contain URIs, labels or a literal (like a boolean value). Matching  URIs and labels can be a particularly difficult task for a question answering benchmarking framework.
\item \emph{C2KB}: This sub-task aims to identify all relevant resources for the given question. It is explained in more detail in Section~\ref{42-sec:extraction}.
\item \emph{P2KB}: This sub-task is similar to C2KB but focusses on the identification of properties $P$ that are relevant for the given question.
\item \emph{RE2KB}: This sub-task evaluates triples that the question answering system extracted from the given search query. The expected answer comprises triples that are needed to build the SPARQL query for answering the question. Note that these triples can contain resources, literals or variables. Two triples are the same if their three elements---subject, predicate and object---are the same. If the triples comprise variables, they must be at the same position while the label of the variable is ignored.
\item \emph{AT}: This sub task identifies the answer type $\alpha$ of the given question. The answer types $\mathbb A$ are defined by the benchmark dataset. In~\cite{qald6}, the authors define 5 different answer types: \texttt{date}, \texttt{number}, \texttt{string}, \texttt{boolean} and \texttt{resource}, where a resource can be a single URI or a set of URIs.
\item \emph{AIT2KB}: This sub-task aims to measure whether the benchmarked question answering system is able to identify the correct type(s) of the expected resources. If the answer does not contain any resources, the set of types $\mathbf{T}$ is expected to be empty. Similar to the ERec task in Section~\ref{42-sec:extraction}, a set of entity types $\mathbb T$ is used as background data for this task.
\end{itemize}
The tasks and their formal descriptions are summarized in Table~\ref{42-tab:QATasks}. The key performance indicators used for the different tasks are the Macro and Micro variants of Precision, Recall and F1-measure. The efficiency of the systems is evaluated by measuring the runtime a system needs to answer a query~\cite{usbeck2019gerbilQA}.

\begin{table*}
\centering
\caption{Summary of the question answering tasks.
}
\begin{tabular}{@{}lccc@{}}
\toprule
\multicolumn{1}{c}{\textbf{Task}} & \textbf{Input} $i_j$ & \textbf{Output} $e_j$ & \textbf{Background data} $B$ \\
\midrule
QA & $q$ & $A \subset K$ & $K$ \\
C2KB & $q$ & $U \subset K$ & $K$ \\
P2KB & $q$ & $P \subset K$ & $K$ \\
RE2KB & $q$ & $\{(s,p,o)|s,p,o \in K \cup V\}$ & $K$ \\
AT & $q$ & $\alpha \in \mathbb{A}$ & $\mathbb{A}$ \\
AIT2KB & $q$ & $\mathbf{T} \subset \mathbb{T}$ & $\mathbb{T}$ \\
\bottomrule
\end{tabular}
\label{42-tab:QATasks}
\end{table*}

Table~\ref{42-tab:datasetsQA} lists datasets based on DBpedia and their features. It is clear that most of the datasets originate from the QALD challenge and contain a small number of questions. This mainly results from the high costs of the manual curation. At the same time, evolving knowledge bases cause a similar problem of outdated datasets, as described in Section~\ref{42-sec:extractionDatasets}. Questions that have been answered on previous versions of a knowledge base might not be answerable on new versions and vice versa. Additionally, the URIs of the resources listed as answers might be different in new versions. A step towards automating the creation of questions is the LC-QuAD dataset~\cite{priyansh2017LCQuad}. The tool for creating this dataset uses SPARQL templates to generate SPARQL queries and their corresponding questions. However, human annotators are still necessary to check the questions and manually repair faulty questions.

\begin{table*}[htb!]
\centering
\caption{Example datasets for benchmarking question answering systems and their features~\cite{priyansh2017LCQuad,usbeck2019gerbilQA}.}
\label{42-tab:datasetsQA}
\begin{tabular}{@{}lp{1cm}rp{1cm}l@{}}
\toprule
Dataset                    & \multicolumn{3}{c}{\#Questions} & Knowledge Base      \\ \midrule
LC-QuAD					   && 5000        && DBpedia 2016-04\\
NLQ shared task 1          && 39          && DBpedia 2015-04                       \\
QALD1\_Test\_dbpedia       && 50          && DBpedia 3.6                           \\
QALD1\_Train\_dbpedia      && 50          && DBpedia 3.6                           \\
QALD1\_Test\_musicbrainz   && 50          && MusicBrainz (dump 2011)\\
QALD1\_Train\_musicbrainz  && 50          && MusicBrainz (dump 2011)               \\
QALD2\_Test\_dbpedia       && 99          && DBpedia 3.7                           \\
QALD2\_Train\_dbpedia      && 100         && DBpedia 3.7                           \\
QALD3\_Test\_dbpedia       && 99          && DBpedia 3.8                           \\
QALD3\_Train\_dbpedia      && 100         && DBpedia 3.8                           \\
QALD3\_Test\_esdbpedia     && 50          && DBpedia 3.8 es                        \\
QALD3\_Train\_esdbpedia    && 50          && DBpedia 3.8 es                        \\
QALD4\_Test\_Hybrid        && 10          && DBpedia 3.9 + long abstracts          \\
QALD4\_Train\_Hybrid       && 25          && DBpedia 3.9 + long abstracts          \\
QALD4\_Test\_Multilingual  && 50          && DBpedia 3.9                           \\
QALD4\_Train\_Multilingual && 200         && DBpedia 3.9                           \\
QALD5\_Test\_Hybrid        && 10          && DBpedia 2014 + long abstracts         \\
QALD5\_Train\_Hybrid       && 40          && DBpedia 2014 + long abstracts         \\
QALD5\_Test\_Multilingual  && 49          && DBpedia 2014                          \\
QALD5\_Train\_Multilingual && 300         && DBpedia 2014                          \\
QALD6\_Train\_Hybrid       && 49          && DBpedia 2015-10 + long abstracts      \\
QALD6\_Train\_Multilingual && 333         && DBpedia 2015-10                       \\ 
\bottomrule
\end{tabular}
\end{table*}

\section{Benchmarking platforms}
\label{42-sec:platforms}

During recent years, several benchmarking platforms have been developed for a single, a subset or all steps of the \gls{ld} lifecycle. In this Section, we give a brief overview of some of these platforms.

\subsection{BAT}

The BAT-framework~\cite{cornolti2013bat} is a one of the first benchmarking frameworks developed for the extraction step of the \gls{ld} lifecycle. Its aim is to facilitate the benchmarking of named entity recognition, named entity disambiguation and concept tagging approaches. BAT compares seven existing entity annotation approaches using Wikipedia as reference.
Moreover, it defines six different task types. Three of these tasks---D2KB, C2KB and A2KB---are described in in Section~\ref{42-sec:extraction}. The other three tasks are mainly extensions of A2KB and C2KB. Sa2KB is an extension of A2KB that accepts an additional confidence score, which is taken into account during the evaluation. Sc2KB is the same extension for the C2KB task. Rc2KB is very similar but expects a ranking of the assigned concepts instead of an explicit score. Additionally, the authors propose five different matchings for the six tasks, i.e., ways how the expected markings and the markings of the system are matched. The framework offers the calculation of six evaluation measures and provides adapters for five datasets.

\subsection{GERBIL}
\label{42-sec:gerbil}

\begin{figure*}
\centering
\includegraphics[width=0.8\linewidth]{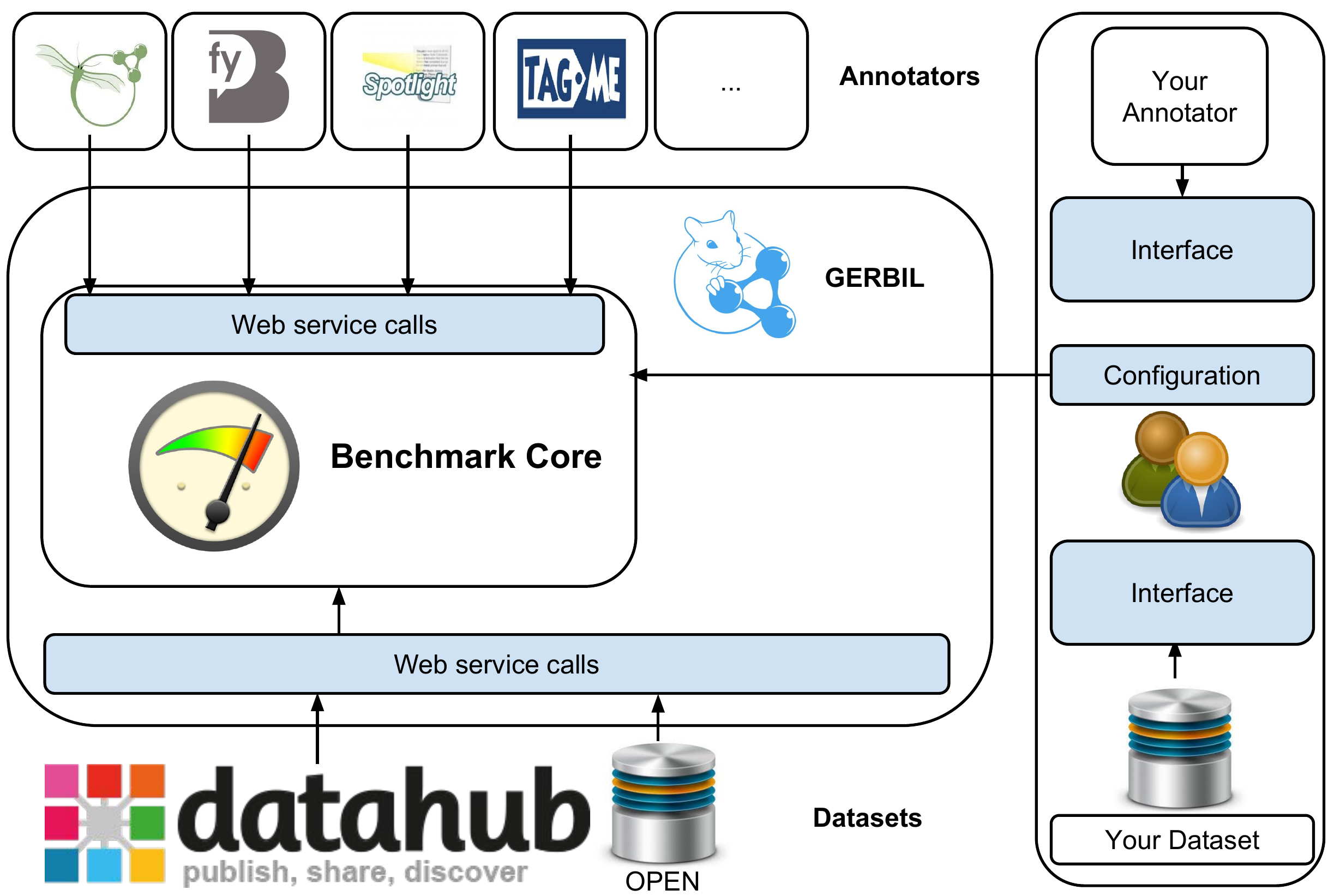}
\caption{Overview of GERBIL's abstract architecture~\cite{usbeck2015gerbil}. Interfaces to users and providers of datasets and annotators are marked in blue.}
\label{42-fig:gerbil}
\end{figure*}

GEBRIL~\cite{usbeck2015gerbil,roeder2018gerbil} is an effort of the knowledge extraction community to enhance the evaluation of knowledge extraction systems. Based on the central idea of the BAT framework (to create a single evaluation framework that eases the comparison and enables the repeatability of experiments),  GERBIL goes beyond these concepts in several ways. Figure~\ref{42-fig:gerbil} shows the simplified architecture of GERBIL. By offering the benchmarking of knowledge extraction systems via web service API calls, as well as the addition of user defined web service URLs and datasets, GERBIL enables users to add their own systems and datasets in an easy way. Additionally, the GERBIL community runs an instance of the framework that can be used for free.\footnote{\url{http://w3id.org/gerbil}} In the sense of the FAIR data principles~\cite{FAIR}, GERBIL provides persistent, citable URLs for executed experiments supporting their repeatability. In its latest version, GERBIL comes with system adapters for more than 20 annotation systems, adapters for the datasets listed in Table~\ref{42-tab:extractionDatasets} and supports the extraction tasks listed in Table~\ref{42-tab:extractionTasks}.\footnote{\url{https://dice-research.org/GERBIL}}
Additionally, GERBIL offers several measures beyond Precision, Recall and F1-measure for a fine grained analysis of evaluation results. These additional measures only take emerging entities or entities from the given knowledge base into account. Another important feature is the identification of outdated URIs. If it is not possible to retrieve the new URI for an entity (e.g., because such a URI does not exist) it is handled like an emerging entity. In a similar way, GERBIL supports the retrieval of \texttt{owl:sameAs} links between entities returned by the system and entities expected by the dataset's gold standard. This enables the benchmarking of systems that do not use the knowledge base the benchmark dataset is relying on.

The success of the GERBIL framework led to its deployment in other areas. GERBIL QA transfers the concept of a benchmarking platform based on web services into the area of question answering~\cite{usbeck2019gerbilQA}.\footnote{\url{http://w3id.org/gerbil/qa}} It supports the evaluation of question answering systems for all tasks listed in Table~\ref{42-tab:QATasks}. It uses the same retrieval mechanism for \texttt{owl:sameAs} links. In addition to that, it uses complex matching algorithms to enable the benchmarking of question answering systems that return labels of entities instead of the entity URIs. 

GERBIL KBC is developed to support the evaluation of knowledge base curation systems---mainly fact validation systems.\footnote{\url{http://w3id.org/gerbil/kbc}} It has been used as the main evaluation tool for the Semantic Web Challenges 2017--2019 and uses Precision, Recall, F1-measure and ROC-AUC as metrics to evaluate the performance of fact validation approaches. Additionally, it supports the organisation of challenges by offering leaderboards including visualisations of the ROC curves of the best performing systems.

\subsection{Iguana}
\label{42-sec:iguana}

\begin{figure*}
\centering
	\includegraphics[width=0.8\textwidth]{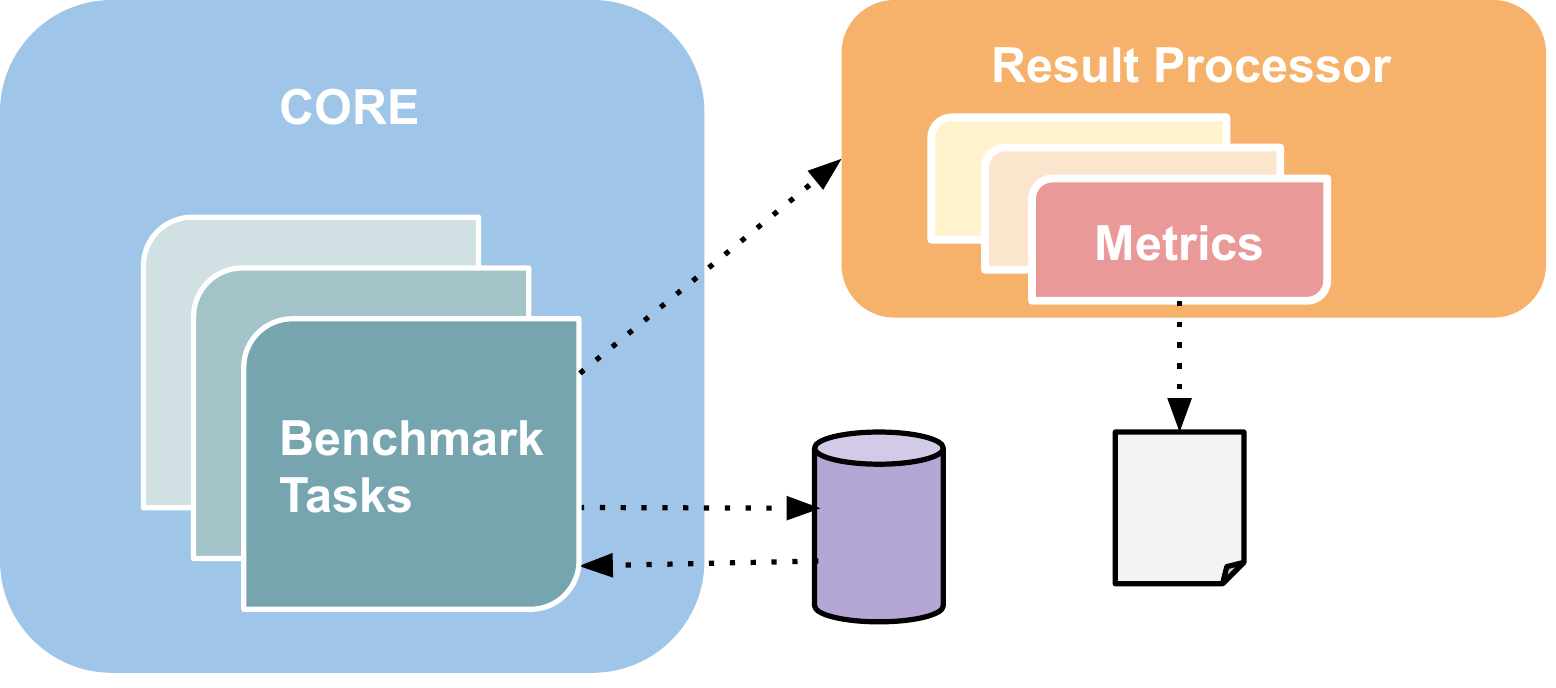}
	\caption{Architecture of Iguana}
	\label{42-fig:iguana}
\end{figure*}

Iguana~\cite{iswc_iguana} is a SPARQL benchmark execution framework that tackles the problem of a fair, comparable and realistic SPARQL benchmark execution.\footnote{\url{https://github.com/dice-group/iguana}} Figure \ref{42-fig:iguana} shows the architecture of Iguana. It consists of a \emph{core} that executes the benchmark queries against the benchmarked triplestores and the \emph{result processor} that calculates metric results.

\subsubsection{Components}

\paragraph{Core.} 
The core uses a given configuration and executes the benchmark accordingly against several provided triplestores. 
To tackle the problem of a realistic benchmark, Iguana implements a highly configurable stresstest, which consists of several user pools that in themselves comprise several threads, representing simulated users. 
Each user pool consists of either SPARQL queries or SPARQL updates that are executed by its threads against the benchmarked triplestore.
Iguana supports SPARQL 1.1 queries as well as SPARQL pattern queries as seen in the DBpedia SPARQL Benchmark \cite{dbpsb2011}. 
Furthermore it supports SPARQL update queries as well as updating Triple files directly.
Each of the parallel-running threads starts at a different query, iterating repeatedly over the given set of queries until either a certain time limit is reached or a certain amount of queries were executed. 
For each query execution the stresstest checks if the query succeeded or failed, and measures the time the triplestore took to answer the query.
This information is forwarded to the result processor.

\paragraph{esult Processor.}

The result processor collects the execution results and calculates pre defined metrics. In its latest version, Iguana supports the following metrics:
\begin{itemize}
	\item Queries Per Second (QPS) counts how often a single query (or a set of queries) can be answered per second.
	\item Query Mixes Per Hour (QMPH) measures how often one query benchmark set can be answered by the triplestore per hour.
	\item Number of Queries Per Hour (NoQPH) counts how many queries can be answered by the triplestore per hour.
\end{itemize}
The calculated results will then be stored as RDF in a triplestore and can be queried directly using SPARQL.

\subsubsection{Benchmark Workflow}
The workflow to benchmark a triplestore using Iguana comprises the following seven steps:
\begin{enumerate}
	\item Create benchmark queries
	\item Start triplestore
	\item Load dataset into triplestore
	\item Create benchmark configuration
	\item Execute configuration using Iguana
	\item Iguana executes benchmark tasks
	\item and provides results
\end{enumerate}

In case the user is not reusing an already existing benchmark dataset, queries have to be created. After that, the triplestore can be started and the dataset can be loaded into the store. A benchmark configuration has to be created containing information like (1) the address of the triplestore, (2) the queries that should be used, (3) the number of user pools and (4) the runtime of the test. Iguana will execute the stresstest as configured flooding the triplestore with requests using the provided queries. After the benchmark finishes, the detailed benchmarking results are available as RDF.

\subsection{SEALS}
\label{42-sec:seals}

The SEALS platform~\cite{seals2012} is one of the major results of the Semantic Evaluation At Large Scale (SEALS) project.\footnote{\url{http://www.seals-project.eu/}} It offers the benchmarking of \gls{ld} systems in different areas like ontology reasoning and ontology matching . It enables the support of different benchmarks by offering a Web Service Business Process Execution Language (WSBPEL)~\cite{WSBPEL2007} interface. This interface can be used to write scripts covering the entire lifecycle of a single evaluation. The platform aims to support evaluation campaigns and has been used in several campaigns during recent years~\cite{Garcia-Castro2011}.

\subsection{HOBBIT}
\label{42-sec:hobbit}
HOBBIT~\cite{roeder2019hobbit} is the first holistic Big Linked Data benchmarking platform. The platform is available as open-source project and as an online platform.\footnote{\url{https://github.com/hobbit-project/platform}, \url{https://master.project-hobbit.eu}} Compared to the previously mentioned benchmarking platforms, the HOBBIT platform offers benchmarks for all steps of the \gls{ld} lifecycle that can be benchmarked automatically. For example, it contains all benchmarks the previously described GERBIL platforms implement. Additionally, it ensures the comparability of benchmark results by executing the benchmarked systems in a controlled environment using the Docker container technology.\footnote{\url{https://docker.com}} The same technology supports the benchmarking of distributed systems and the execution of distributed benchmarks. The latter is necessary to be able to generate a sufficient amount of data to evaluate the scalability of the benchmarked system.

\begin{figure*}
\centering
\includegraphics[scale=0.5]{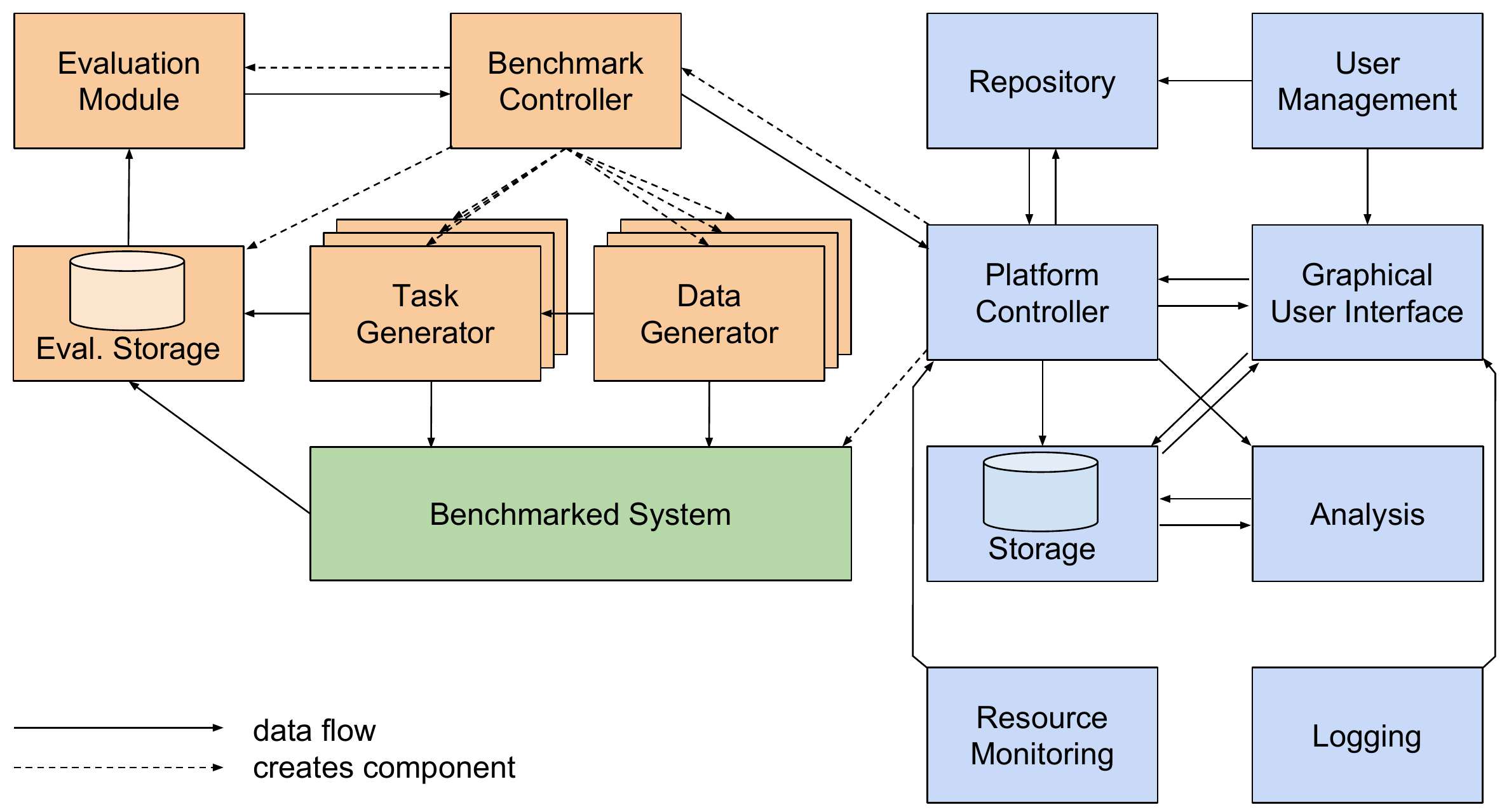}
\caption{The components of the HOBBIT platform~\cite{roeder2019hobbit}.}
\label{42-fig:hobbit-components}
\end{figure*}

Figure~\ref{42-fig:hobbit-components} shows the components of the HOBBIT platform. The blue boxes represent the components of the platform. One of them is the graphical user interface allowing the interaction of the user with the platform. A user management ensures that not all users have access to all information of the platform. This is necessary for the support of challenges where the challenge organiser has to setup the configuration of the challenge benchmark in a secret way. Otherwise, participants could adjust their solutions to the configuration.
The storage contains all evaluation results as triples and offers a public read-only SPARQL interface. The analysis components offers the analysing of evaluation results over multiple experiments. This can lead to helpful insights regarding strengths and weaknesses of benchmarked systems. The resource monitoring enables the benchmark to take the amount of resources used by the benchmarked system (e.g., CPU usage) into account. The logging component gives benchmark and system developers access to the log messages of their components. 

Although the HOBBIT platform supports nearly every benchmark implementation, the authors of~\cite{roeder2019hobbit} suggest a general structure for a benchmark which separates it into a controller that orchestrates the benchmark components, a set of data generators responsible for providing the necessary datasets, a set of task generators creating the tasks the benchmarked systems has to fulfill, an evaluation storage that stores the expected answers and the responses generated by the benchmarked system and an evaluation module that implements the metrics the benchmark uses. A detailed description of the components and the workflow is given in~\cite{roeder2019hobbit}.

\section{Conclusion}
\label{42-sec:conclusion}

In this chapter, we presented an overview of benchmarking approaches for the different steps of the Linked Data lifecycle. These benchmarks are important for the usage and further development of \gls{kg}-based systems since they enable a comparable evaluation of the system's performance. Hence, for the development of explainable artificial intelligence based on \glspl{kg} these benchmarks will play a key role to identify the best KG-based systemsfor interacting with the \gls{kg}. In future works, these general benchmarks might be adapted towards special requirements explainable artificial intelligence approaches might raise with respect to their underlying \gls{kg}-based systems.

\bibliographystyle{acl_natbib}
\bibliography{42-roeder}

\end{document}